\documentclass[final,5p,times,twocolumn]{elsarticle}

\usepackage{amsmath,amssymb,amsfonts}
\usepackage{algorithmic}
\usepackage{graphicx}
\usepackage{textcomp}
\usepackage{soul}
\usepackage{booktabs}
\usepackage{makecell}
\usepackage{colortbl}
\usepackage{multirow}
\usepackage{tcolorbox}
\usepackage{subcaption}
\usepackage[bookmarks=false]{hyperref}
\usepackage{url}
\usepackage{xcolor}
\usepackage{balance}
\usepackage[textsize=scriptsize]{todonotes}

\tcbuselibrary{skins}
\NewTotalTColorBox[auto counter]{\Definition}{ +m }{
    notitle,
    colback=blue!3!white,
    frame hidden,
    boxrule=0pt,
    enhanced,
    sharp corners,
    borderline west={4pt}{0pt}{blue!20!white},
     boxsep=2pt,  
    left=7pt,    
    right=2pt,   
    top=2pt,     
    bottom=2pt,  
}{
    \textcolor{green!50!black}{
        \sffamily
    }%
    #1
}

\def\BibTeX{{\rm B\kern-.05em{\sc i\kern-.025em b}\kern-.08em
    T\kern-.1667em\lower.7ex\hbox{E}\kern-.125emX}}

\newcommand{\toolname}[1]{\emph{GenioSim}}

\begin{document}

\begin{frontmatter}

\title{\toolname{}: A Novel Simulation Platform for Edge Computing over Optical Networks}

\author[inst1]{Carmine Cesarano}
\author[inst2]{Alessio Foggia}
\author[inst3]{Roberto Natella}

\address[inst1]{KTH Royal Institution of Technology, Stockholm, Sweden}
\address[inst2]{University of Naples Federico II, Naples, Italy}
\address[inst3]{Gran Sasso Science Institute, L'Aquila, Italy}

\begin{abstract}
The convergence of Passive Optical Networks (PONs) and edge computing creates new opportunities: Optical Line Terminals (OLTs) and Optical Network Terminals (ONTs) can be repurposed as low-latency edge compute nodes for offloading workloads. 
However, exploring such design options early in the development cycle is costly and time-consuming, as prototyping requires specialized hardware and realistic traffic conditions.
Simulation becomes essential, yet current tools are unable to accurately model this emerging class of systems. 
To address these gaps, we introduce \toolname{}, a simulation platform for hierarchical PON-enabled edge infrastructures. It models OLTs and ONTs with realistic PON behavior, supports hybrid container- and VM-based virtualization, and provides multiple service and execution models. These capabilities enable the evaluation of resource management policies under complex, heterogeneous conditions.
We present experiments in the context of use cases of industrial relevance, to show \toolname{} can provide insights for capacity planning and for the choice of policies for container placement and task offloading in PON-enabled edge infrastructures.
\end{abstract}

\begin{keyword}
Simulation, Modeling, Edge Computing, Passive Optical Networks
\end{keyword}

\end{frontmatter}

\section{Introduction}
Passive Optical Networks (PONs) have become the \emph{de facto} standard for fiber-to-the-home (FTTH) access, with widespread deployment and continued global growth~\cite{broadband_news}. Optical Line Terminals (OLTs), located in central offices of service providers, serve as aggregation points for multiple user premises over optical fibers. These terminals are typically backed up by commercial off-the-shelf (COTS) components, including x86 CPUs and high-capacity storage hardware, and open-source software (OSS), including operating systems, orchestrators, and virtualization technologies, to run network management functions, such as dynamic bandwidth allocation and control-plane signaling, following the Network Function Virtualization (NFV) paradigm~\cite{survey_on_nfv}. 

This trend has sparked growing interest in repurposing the existing PON backend infrastructure to support edge computing workloads. 
The edge computing paradigm has gained traction as a promising approach to meet increasingly stringent latency and bandwidth requirements, by leveraging computational resources closer to end-users~\cite{KHAN2019219}. 
Specifically, co-locating edge functions alongside the OLT within central offices is now being actively explored by both industry and academia~\cite{telefonica_whitepaper, tim_edge_cloud, cesarano2025genio}, aiming to achieve high-throughput and ultra-low-latency processing capabilities.

Consequently, there is a growing need for flexible solutions that design and validate these systems under realistic workloads, resource constraints, and dynamic network conditions. In particular, \emph{simulation} is a key pillar for the analysis of computer systems, since it enables the evaluation of complex systems before they are physically built, saving time and resources, and potentially preventing costly errors~\cite{fi11030055}. Indeed, physical PON testbeds are expensive to build, requiring specialized hardware, complex middleware and software integration, and configuration efforts across multiple layers of the network stack. These high barriers to experimentation limit the ability to rapidly prototype and explore alternative designs, deployment strategies, and orchestration algorithms. 
Therefore, simulation tools are essential for capacity planning and performance evaluation before committing to costly physical deployments.

Despite the availability of edge computing simulators, a critical research gap remains in their ability to model heterogeneous infrastructures, particularly those that incorporate access technologies such as PONs. Existing tools~\cite{sonmez2018edgecloudsim, mechalikh2019pureedgesim, souza2023edgesimpy} have made significant progress by introducing several features, such as user mobility modeling, modular service composition, and energy-aware resource simulation. However, they do not cover well the following three key areas. First, (1) current simulators cannot model complex, hierarchical network topologies, in particular the multi-level infrastructure of PONs, including splitters, Optical Network Terminals (ONTs), and the co-location of compute and network functions within central offices~\cite{DIAS2023103191}. Second, (2) there is limited support for hybrid virtualization environments: while some tools model workload execution in \emph{virtual machines} (VMs), they do not reflect setups that combine VMs for the management of physical resources of the PON infrastructure (e.g., to isolate network management functions from edge workloads) and \emph{containers}, for lightweight isolation between edge applications~\cite{lightweight_virtualization}. Third, (3) most simulators focus primarily on \emph{task offloading}, where clients upload and execute individual tasks on edge computing servers: however, more complex business service models deploy long-running applications on edge servers (\emph{service placement}), to reduce response latency and share resources among multiple service requests \cite{qian2019privacy,he2018s,guo2023edge,chen2024joint}.
These limitations hinder the ability of existing simulators to experiment with realistic hierarchical orchestration strategies in PON-enabled infrastructures.

To address these limitations, we introduce \textbf{\toolname{}}, a novel simulation platform. 
\toolname{} is inspired by \textit{GENIO} (\emph{Edge cloud platform enabled by a new intelligent OLT for GPON networks}), an industrial R\&D project on a new orchestration platform for edge workloads in PON networks~\cite{cesarano2025genio}. 
\toolname{} supports realistic PON topologies, including ONTs, OLTs, and optical fiber links, as well as their hierarchical relationships.  
Moreover, \toolname{} includes native support for hybrid virtualization, enabling simulations of both containers and virtual machines, and introduces a decoupled execution model in which applications (as containers) and user requests (as tasks) follow distinct workflows. Unlike existing simulators, \toolname{} enables the joint evaluation of \emph{service placement} and \emph{task offloading} policies over hierarchical PON infrastructures, under heterogeneous resources and multi-layer orchestration constraints.
These extensions enable the analysis of adaptive orchestration strategies and workload distribution across realistic PON-enabled edge computing scenarios.

To demonstrate the capabilities of \toolname{}, we present a set of experiments in the context of use cases of industrial interest from the GENIO project. The experiments explore two representative uses of \toolname{}: (i) \emph{capacity planning}, where \toolname{} is used to determine the minimum computational resources required in OLTs and ONTs to satisfy application-specific latency SLOs; and (ii) \emph{orchestration policy evaluation}, where \toolname{} evaluates how different container placement and task offloading strategies interact with the PON hierarchy and heterogeneous edge resources. Overall, the results show that \toolname{} is not only able to model these infrastructures, but also to expose non-trivial trade-offs in capacity sizing, workload distribution, and policy selection. In particular, the experiments show that far-edge resources improve performance only when coordinated through resource-aware and delay-aware orchestration policies, whereas naive proximity-driven choices may degrade both latency and task success rate. To foster reproducibility and broader adoption, we release \toolname{} as open-source software\footnote{https://github.com/dessertlab/GenioSim}.
The main contributions of the paper are summarized as follows:

\begin{itemize}
\setlength{\itemsep}{0pt}
\setlength{\parskip}{0pt}

    \item We present \toolname{}, a simulation platform for PON-enabled edge computing infrastructures with explicit modeling of hierarchical PON topologies.

    \item We implement in \toolname{} several strategies for container placement and task offloading, enabling the systematic analysis of orchestration policies for PON-enabled edge computing.

    \item We apply \toolname{} for both capacity planning and for the evaluation of orchestration policies on use cases derived from an industrial R\&D project.
\end{itemize}

The remainder of the paper is organized as follows. Section~\ref{sec:background} provides background on PONs, resource orchestration in edge computing environments, and the GENIO project. Section~\ref{sec:related} discusses related work. Section~\ref{sec:design} presents the design of \toolname{}. Section~\ref{sec:exp} describes the experimental evaluation. Section~\ref{sec:scalability} analyzes the scalability and performance of the simulator. Section~\ref{sec:discussion} discusses limitations of the current model and outlines directions for future work and Section~\ref{sec:conclusion} concludes the paper.

\section{Background}
\label{sec:background}

\subsection{Passive Optical Networks}
Passive Optical Networks (PONs) are widely deployed access technologies for fiber-to-the-home (FTTH) connectivity. A typical PON topology consists of an \emph{Optical Line Terminal} (OLT) located in the service provider's central office, multiple \emph{Optical Network Terminals} (ONTs) at user premises, and passive optical splitters that connect them. The OLT acts as an aggregation point, concentrating user traffic and interfacing with metro or core networks. Although originally designed for broadband access, the hierarchical structure of PON, where lightweight ONTs forward traffic to OLTs, makes them a promising substrate for distributed computing. In particular, OLTs are increasingly equipped with commercial off-the-shelf (COTS) servers and storage, enabling their reuse as edge computing nodes co-located with access infrastructure.

\subsection{Edge Computing and Virtualization Models}
Edge computing extends cloud capabilities by deploying resources closer to end-users, thereby reducing latency and the load on backhaul. In practice, service providers rely on virtualization technologies to manage these resources. \emph{Virtual Machines} (VMs) offer strong isolation and are typically used to host network management functions and virtualized infrastructure services. \emph{Containers}, in contrast, provide lightweight isolation with faster startup times, making them suitable for edge applications that require dynamic and fine-grained scaling. Hybrid deployments are increasingly common in telco environments, where VMs and containers coexist: VMs host critical infrastructure services, while containers support user-facing workloads. Modeling this coexistence is essential for a realistic evaluation of edge computing platforms.

\subsection{Service Placement and Task Offloading}
Resource orchestration in edge infrastructures involves two complementary processes. \emph{Service placement} determines where long-lived applications (e.g., containers) are deployed across available nodes. Once deployed, multiple user requests can be served by the same application instance, balancing resource costs. \emph{Task offloading}, in contrast, occurs at runtime and assigns individual user tasks to one of the available service instances, based on factors such as latency, load, or network conditions. Decoupling placement from offloading enables adaptive orchestration strategies, particularly in hierarchical infrastructures where services may span cloud, edge, and far edge. This distinction is critical for supporting realistic workload beyond the simplified task-execution model of existing simulators. 

\begin{figure}[t]
  \includegraphics[width=\linewidth]{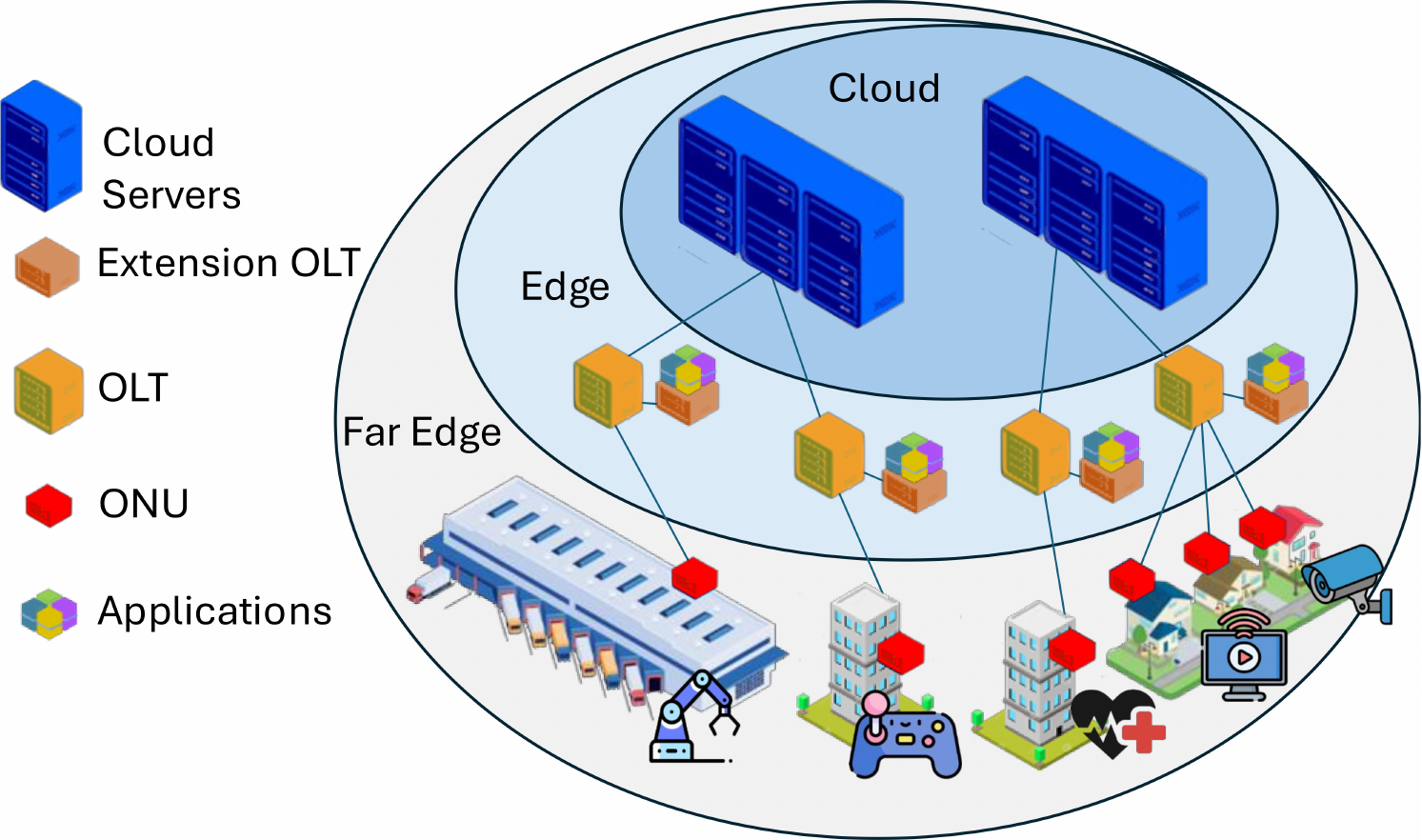}
  \caption{GENIO Project deployment.}
  \label{fig:GENIO_deployment}
\end{figure}

\subsection{Genio Project}
The GENIO project develops an industrial edge-computing platform integrated with PON infrastructures, where computation is distributed across three hierarchical tiers, ONTs at the far edge, OLTs at the edge tier, and the cloud, as shown in Figure~\ref{fig:GENIO_deployment}. ONTs are lightweight devices deployed in customer premises, capable of executing ultra-low-latency tasks or forwarding them upstream. OLTs, traditionally used for PON management, are repurposed as compute-capable edge hubs based on x86 COTS hardware and extended with an SDN stack to dynamically manage access-network traffic and multi-tenant resources. The cloud tier provide high-capacity computing and hosts the global orchestrator responsible for service placement, resource management, and cross-OLT coordination.

A defining characteristic of GENIO is the coexistence of hybrid virtualization, where VM-based isolation (managed via KVM and Proxmox) is complemented by container-based deployments for business applications (managed via Kubernetes). This mixed model reflects real operational constraints: infrastructure services require strong VM isolation, while tenant applications demand lightweight, cloud-native deployment models. Consequently, GENIO applications may run in dedicated VMs, containers inside VMs, or shared container pools, making hybrid virtualization essential for any simulator aiming to evaluate realistic placement and offloading decisions. 

GENIO targets industrial and telco-grade workloads, including IoT analytics, computer-vision pipelines, smart-building monitoring, ML inference services, and multimedia streaming. These applications exhibit heterogeneous compute demand, latency SLOs, and user access patterns, often with bursty workloads and multi-tenant sharing. The platform adopts a service-oriented model in which long-running containerized services are deployed once an then repeatedly invoked by multiple users, a key difference from simulators where tasks implicitly instantiate their own execution environment.

Another crucial architectural trait is hierarchical orchestration: the cloud decides where services (containers) are placed across ONTs and OLTs, while runtime task routing uses DNS-based brokers deployed at OLTs to direct user requests to the nearest or least-loaded service replica. This decentralized routing layer is necessary for telco-scale deployments and must therefore be explicitly modeled in simulation.

Together, these characteristics, and the lack of existing simulators that support them, directly motivate the design choices behind \toolname{}.

\begin{figure}[t]
  \includegraphics[width=\linewidth]{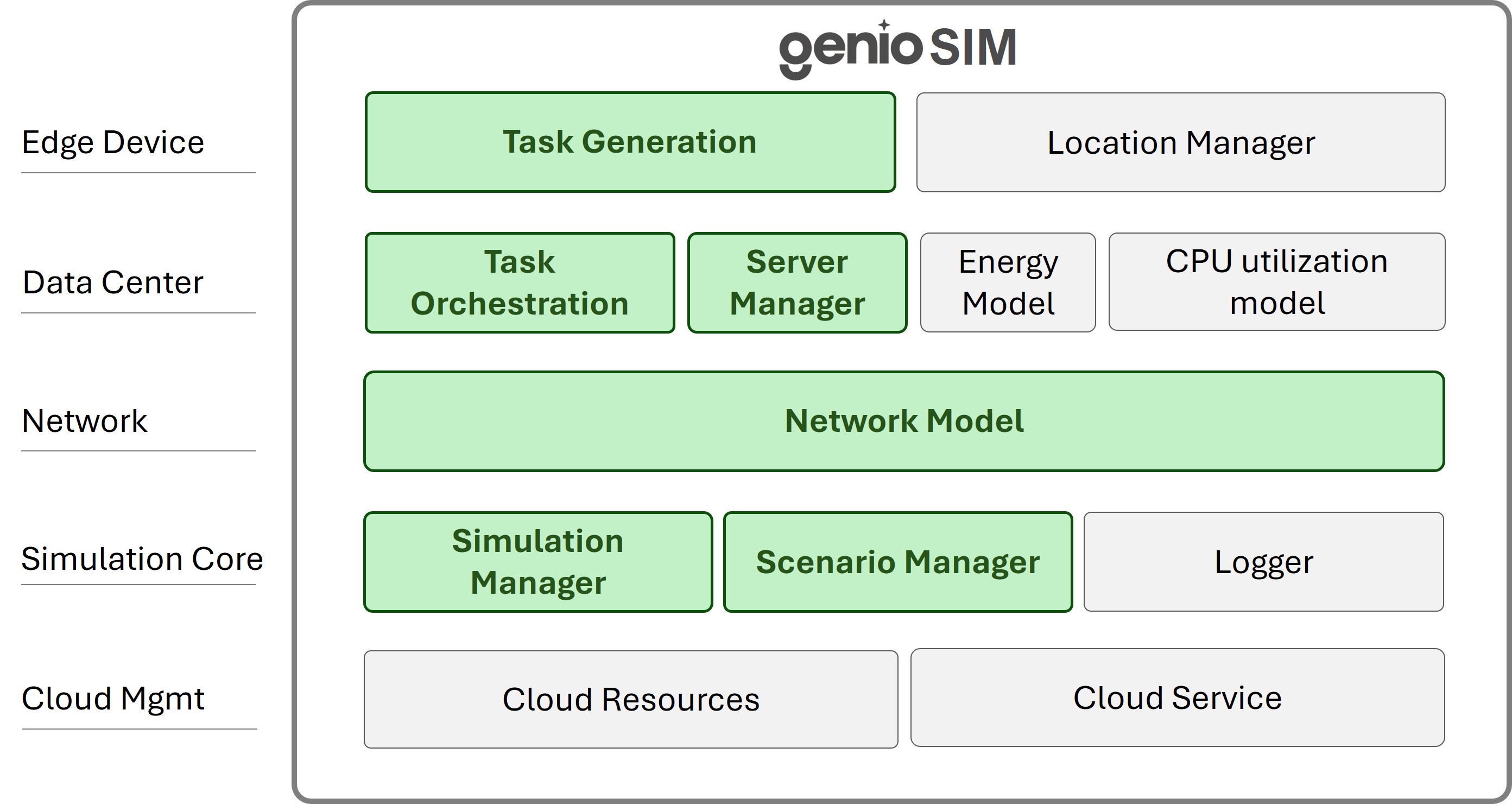}
  \caption{Architectural overview of \toolname{}. The modules highlighted in green represent the areas of enhancement with respect to the state-of-the-art.}
  \label{fig:architecture}
\end{figure}

\section{Related Work}
\label{sec:related}

\paragraph{Cloud and Edge Computing Simulators}
Simulation is widely used to evaluate edge infrastructure and orchestration strategies before deployment, complementing costly and less flexible testbeds. CloudSim and derivatives focus on data-center resource management with abstracted networks and no edge devices~\cite{cloudsim}. Edge-oriented simulators, including EdgeSimPy, EdgeCloudSim, iFogSim/iFogSim2, FogNetSim++, YAFS, LEAF, ENIGMA, MyIFogSim, and PureEdgeSim extend this paradigm to three-tier (cloud, edge, device) settings ans support task-based workloads and placement mechanisms~\cite{souza2023edgesimpy, sonmez2018edgecloudsim, gupta2017ifogsim, mahmud2022ifogsim2, qayyum2018fognetsim, lera2019yafs, wiesner2021leaf, del2023scalable, lopes2017myifogsim, mechalikh2019pureedgesim}. However, these tools rely on simplified resource and network abstractions and primarily target consumer scenarios rather than industrial deployments~\cite{margariti2020modeling, caiazza2021measurement, gomez2024energy, kaftantzis2024exploring, belcastro2025navigating}. PureEdgeSim is most relevant due to its modular architecture~\cite{mechalikh2019pureedgesim}, but its network, virtualization, and orchestration models cannot capture the PON-enabled, cloud-native, multi-tenant settings targeted by GENIO. \toolname{} extends PureEdgeSim to address these gaps.

\paragraph{Network and Access Infrastructure Modeling}
Simulators differ widely in network modeling: FogNetSim++ and YAFS offer flexible topologies~\cite{qayyum2018fognetsim, lera2019yafs}, yet most tools still represent communication through coarse bandwidth, such as delay links without detailed lower-layer or access-technology behavior~\cite{margariti2020modeling, caiazza2021measurement}. This abstraction is problematic for industrial edge computing like GENIO, where PON is central. Existing simulators, including iFogSim, EdgeCloudSim, YAFS, and PureEdgesim, assume that flat networks in which devices directly connect to edge/cloud nodes, omitting ONT-OLT hierarchies and fiber aggregation~\cite{gupta2017ifogsim, sonmez2018edgecloudsim, lera2019yafs, mechalikh2019pureedgesim}. Even energy-focused tools like LEAF lack PON-specific entities~\cite{wiesner2021leaf, gomez2024energy}. 
Research on multi-access edge computing and network softwarization highlight the importance of optical access~\cite{kim2020dynamic, gharbaoui2021resource}, but these works provide algorithms, not reusable simulation environments. No existing simulator explicitly models ONTs/OLTs or PON congestion.
\toolname{} fills this gap by introducing ONTs/OLTs and optical links with configurable latency, bandwidth, and energy.

\paragraph{Virtualization and Cloud-Native Modeling}
With cloud-native architectures becoming standard, containers and Kubernetes-based orchestration dominate modern edge deployments, while hybrid VM+container environments remain common~\cite{moreno2021emulating, bohm2022cloud, adoga2023towards, leivadeas2019vnf}. Existing simulators remain VM-centric: CloudSim and derivatives treat VMs as the main compute unit~\cite{cloudsim}, and EdgeCloudSim and iFogSim variants similarly execute tasks directly on hosts or VMs without explicit container modeling~\cite{sonmez2018edgecloudsim, gupta2017ifogsim, mahmud2022ifogsim2}. Calls for first-class support of containers, VMs, and serverless functions remain largely unmet~\cite{raith2023faas}. 
\toolname{} adds a hybrid virtualization layer where containers run inside VMs at OLTs and ONTs, modeling hypervisor overhead through a new network-level abstraction and configurable XML-based deployment policies.

\paragraph{Orchestration and Request Dispatch}
Edge simulators such as FogNetSim++, iFogSim, and EdgeCloudSim implement placement and scheduling policies accounting for node utilization and application latency~\cite{qayyum2018fognetsim, gupta2017ifogsim, sonmez2018edgecloudsim}, but typically assume centralized controllers and shallow hierarchies~\cite{caiazza2021measurement}. A major limitation is the lack of joint modeling of service placement and request dispatch: placement is studied in isolation from routing, despite their tight coupling~\cite{shen2020ai}. Centralized cloud-native schedulers also conflict with multi-tier edge environments~\cite{hsiao2021optimization, shen2020ai, kim2020dynamic}. Hierarchical orchestration schemes have been proposed~\cite{bartolomeo2207oakestra, gharbaoui2021resource}, but focus on algorithms rather than simulation frameworks. 
\toolname{} introduces hierarchical orchestration with separate placement (\texttt{ContainerOrchestrator}) and task-offloading (\texttt{TaskOrchestrator}) layers, coordinated via DNS-based brokers deployed on OLT servers.

\paragraph{Workload and User Modeling}
Existing tools typically model workload as independent tasks defined by arrival rates rather than long-lived applications~\cite{sonmez2018edgecloudsim, gupta2017ifogsim}. These approaches inadequately capture industrial workloads exhibiting complex, bursty, multi-tenant dynamics~\cite{gomez2024energy, belcastro2025navigating, kaftantzis2024exploring}. 
\toolname{} separates services from users: services are persistent containers with configurable resource and network footprint, while user profiles define heterogeneous access patterns, burstiness, and temporal behavior, enabling realistic multi-tenant and session-based dynamics.

\paragraph{Positioning of \toolname{}}
To sum up, existing simulators lack (i) PON-aware access hierarchies, (ii) hybrid VM-container modeling, (iii) joint placement-offloading designs, an (iv) realistic user-modeling and multi-tenant workload.
\toolname{} extends PureEdgeSim with explicit PON entities, hybrid virtualization with hypervisor-aware overhead, hierarchical orchestration with DNS-based brokers, and decoupled service/user modeling. This integrated design enables realistic evaluation of orchestration strategies in PON-enabled industrial environments.

\section{\toolname{} Design}
\label{sec:design}

\toolname{} has been designed to overcome limitations of existing edge computing simulators. As discussed in the previous section, other simulators cannot accurately represent access technologies and complex orchestration models, thus falling short in modeling the industrial use cases targeted by the GENIO project. To address these gaps, the design of \toolname{} is guided by four requirements:

\begin{enumerate}
\item \textit{Realistic modeling of PON-enabled infrastructures:} capturing the hierarchical structure of PON, including ONTs, OLTs, and fiber optic links.

\item \textit{Support for hybrid virtualization:} enabling coexistence of VMs and containers, reflecting cloud-native deployments.

\item \textit{Hierarchical orchestration:} decoupling high-level service placement from task execution, distributed across cloud, edge, and far-edge tiers.

\item \textit{Realistic workload and user modeling:} supporting long-lived services, multi-tenancy, and heterogeneous user behaviors.
\end{enumerate}

These requirements extend PureEdgeSim into a PON-aware, cloud-native simulation environment that aligns with the needs of the GENIO project, addressing the gaps identified in Section~\ref{sec:related}.

\toolname{} adopts a \emph{system-level simulation approach}, following the design philosophy of other popular simulators such as CloudSim and PureEdgeSim~\cite{cloudsim, mechalikh2019pureedgesim}. Rather than modeling network behavior at the packet level, the simulator focuses on application-level aspects relevant for edge orchestration studies, such as service placement, task offloading, and resource allocation across heterogeneous nodes. This abstraction enables the exploration of orchestration strategies in large-scale PON-enabled edge infrastructures while maintaining practical simulation scalability.

\begin{figure*}[h]
  \centering
  \begin{subfigure}[t]{0.48\linewidth}
    \centering
    \includegraphics[width=\linewidth]{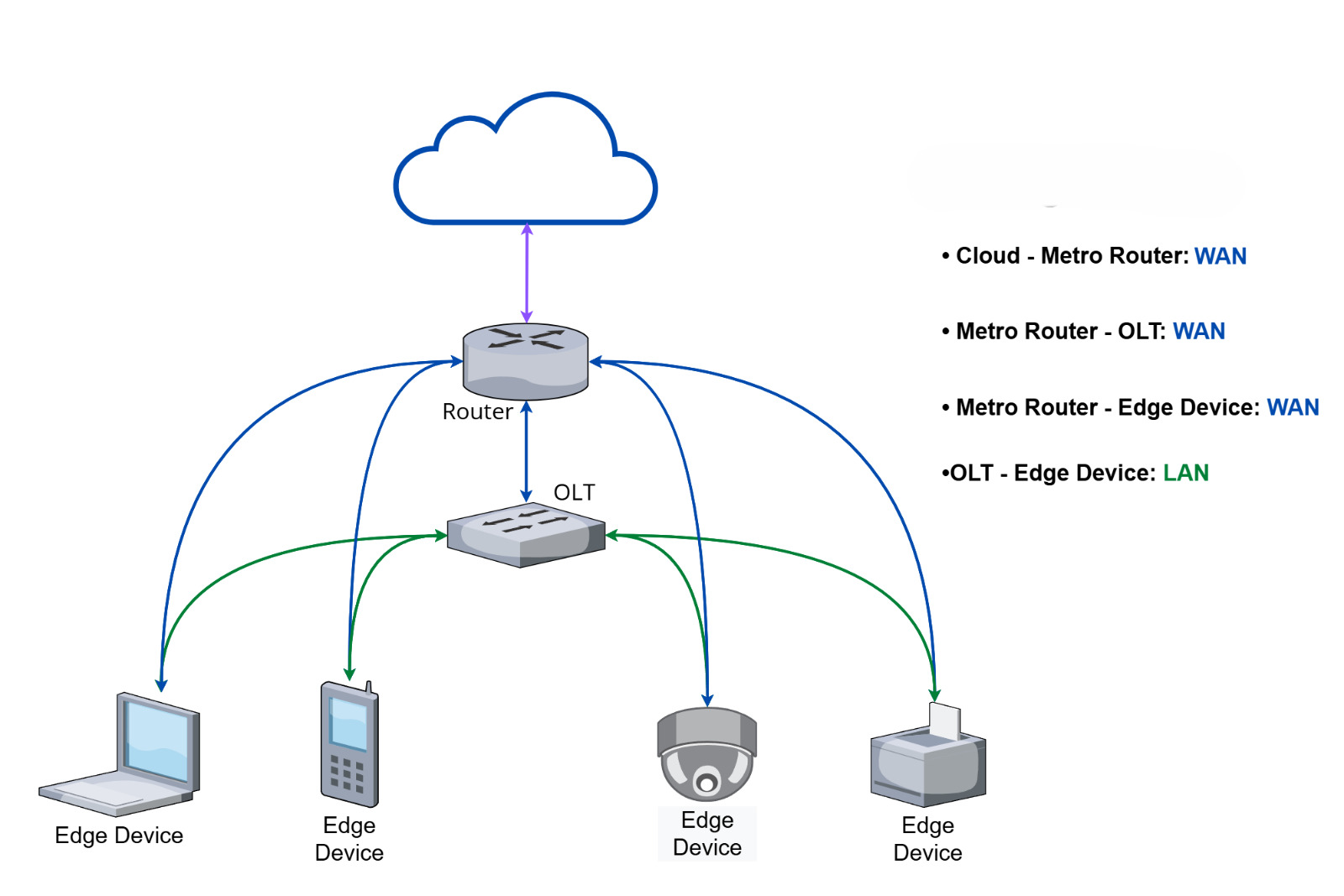}
    \label{fig:network_before}
  \end{subfigure}%
  \hfill
  \begin{subfigure}[t]{0.48\linewidth}
    \centering
    \includegraphics[width=\linewidth]{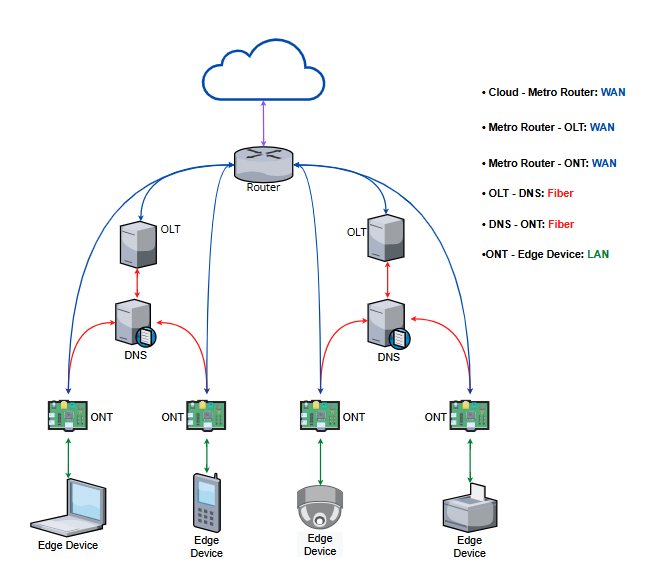}
    \label{fig:network_after}
  \end{subfigure}
  \caption{Comparison of network topologies supported by PureEdgeSim (left) vs. GenioSim (right).}
  \label{fig:network_comparison}
\end{figure*}

\subsection{Architectural Overview}

\toolname{} builds on a modular architecture that separates simulation control, infrastructure modeling, workload generation, networking, orchestration, and scenario configuration.
Figure~\ref{fig:architecture} illustrates the main modules and highlights the components that extend or replace those of earlier simulators such as PureEdgeSim~\cite{mechalikh2019pureedgesim} and CloudSim~\cite{cloudsim}.
While retaining this modular foundation for compatibility and extensibility, \toolname{} introduces a set of targeted enhancements that involve multiple modules.

\paragraph{Simulation Manager}
Handles the global event loop, lifecycle timing, and initialization of containers and tasks. Unlike previous simulators, it supports decoupled service placement and task execution, allowing long-running services to be deployed independently of user-generated requests.

\paragraph{Server Manager}
Models compute resources at the cloud, edge, and far-edge layers. \toolname{} extends this module with explicit PON-aware entities (ONTs, OLTs), hierarchical resource organization, and a hybrid virtualization layer where VMs host containers.

\paragraph{Network Module}
Represents communication across hierarchical PON topologies, including ONT–OLT relationships and optical fiber links. Link models include configurable latency, bandwidth, and energy, enabling realistic bottleneck and congestion behavior.

\paragraph{Task Orchestration Module}
Implements hierarchical orchestration by separating container placement from task offloading. Placement determines where services reside; offloading determines, at runtime, how tasks flow through ONTs, OLTs, and DNS-based brokers.

\paragraph{Scenario Manager and Task Generator}
Provide a high-level configuration interface (via XML) for defining applications, container properties, replica placement, and heterogeneous user behavior. This supports multi-tenant and bursty workloads reflective of real deployments.

\bigbreak
In the following sections, we detail the main characteristics that differentiate \toolname{} from prior simulators.

\subsection{PON-enabled Modeling}
In previous simulators, such as PureEdgeSim, the \texttt{Network Module} models communication as a flat structure, where devices are directly connected to cloud or edge data centers. Bandwidth allocation adapts dynamically to load, but the model lacks hierarchical access topologies as those found in PONs.  

In \toolname{}, the \texttt{Server Manager} and the \texttt{Network Module} explicitly represent the entities of PON infrastructures. Each edge device is paired with an \textit{ONT}, which acts as a lightweight far-edge node. \textit{ONTs} forward traffic to \textit{OLTs}, which aggregate flows and act as edge-level computing hubs. The \texttt{Network Module} includes \textit{optical fiber links} as configurable connections with tunable parameters to represent latency, bandwidth, and energy consumption.
Figure~\ref{fig:network_comparison} shows the differences between PureEdgeSim and \toolname{} with respect to the modeled network topologies.

\toolname{} introduces \textit{brokers}, based on the DNS protocol. These entities act as intermediaries between end-users and services, by routing each user request towards the service instance that will handle it, and balancing the load among instances. These DNS-based brokers do not provide computational resources, but only handle routing and load balancing. They are hosted on OLT servers at the edge layer.

This design mirrors the GENIO platform, where DNS-based routing is used as a lightweight, decentralized, and highly scalable mechanism for directing user requests across the telco edge. DNS provides a fault-tolerant method for mapping service names to dynamically changing service endpoints without relying on centralized controllers or heavy service meshes, scales naturally to multi-OLT deployments, and enables fast traffic redirection under dynamic placement changes, failures, or load variations.

By explicitly modeling in detail how user requests traverse ONTs, OLTs, DNS brokers, and upstream links, \toolname{} supports fine-grained analysis of traffic distribution, dynamic workload redirection, and the interplay between placement decisions and runtime task offloading. This design makes it possible to capture congestion phenomena across the PON infrastructure, including overloads and bottlenecks at ONTs, OLTs, DNS brokers, and fiber access links.

\subsection{Hybrid Virtualization}
\begin{figure}[h]
  \includegraphics[width=\linewidth]{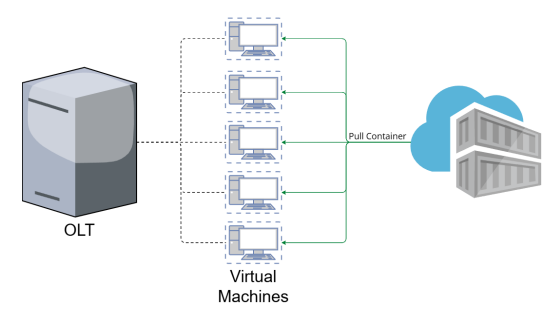}
  \caption{Hybrid virtualization in \toolname{}, where containers are deployed inside VMs hosted on OLTs.}
  \label{fig:virt}
\end{figure}

In PureEdgeSim, the \texttt{Server Manager} models data centers, servers, and VMs, but does not support explicit containerization. Tasks are executed directly on hosts or VMs, limiting the ability to simulate cloud-native deployments. 
\toolname{} provides a \texttt{Server Manager} with a \textit{hybrid virtualization layer}, where containers can be deployed inside VMs hosted on OLTs and ONTs (Figure~\ref{fig:virt}). This layered model mirrors real-world PON-based infrastructures where VMs provide strong isolation for infrastructure services and containers support the lightweight, flexible deployment of edge applications.  

To enable this model, the \texttt{Network Module} introduces a new \textit{hypervisor link} abstraction that connects containers to their hosting VM. This link accounts for configurable latency and energy overheads associated with hypervisor mediation, which were ignored in PureEdgeSim. These costs are explicitly reflected in task execution times, allowing experiments to capture the trade-offs between isolation and performance. 

As a result, \toolname{} supports fine-grained comparison between different deployment choices: pure VM execution, pure container execution, or mixed setups. Configuration files in XML format specify VM capacity, container allocation policies (private vs. shared), and the number of replicas, enabling the exploration of virtualization strategies that affect resource efficiency and QoS.

\subsection{Hierarchical Orchestration}
\label{subsec:orchestration}
One of the requirements in the GENIO project is the ability to support hierarchical orchestration, where long-lived services are deployed independently of the short-lived user requests that consume them. In PureEdgeSim, the \texttt{Task Orchestration Module} couples these two dimensions: tasks are served by instantiating a job on demand on an edge server, thereby conflating service creation with request execution. This prevents modeling realistic cloud-native deployments where services run continuously, are reused by multiple clients, and where task arrivals occur at a high rate. It also contradicts GENIO’s operational model, where services are provisioned once and then shared across many users.

\toolname{} introduces a \textit{hierarchical orchestration model} that explicitly separates service placement from task offloading, aligning the simulator with real-world PON-enabled architectures and GENIO’s service subscription paradigm:

\begin{itemize}
    \item \textit{Service placement:} a global orchestrator determines where each service replica is deployed, across cloud nodes, OLTs, or ONTs. Services are represented as long-lived containers whose lifecycle (deployment, migration, removal) is independent of user activity.
    \item \textit{Task offloading:} at runtime, user requests are routed by DNS-based brokers, which select the most suitable service replica based on current latency, load, and resource availability.
\end{itemize}

\begin{figure}[h]
  \includegraphics[width=\linewidth]{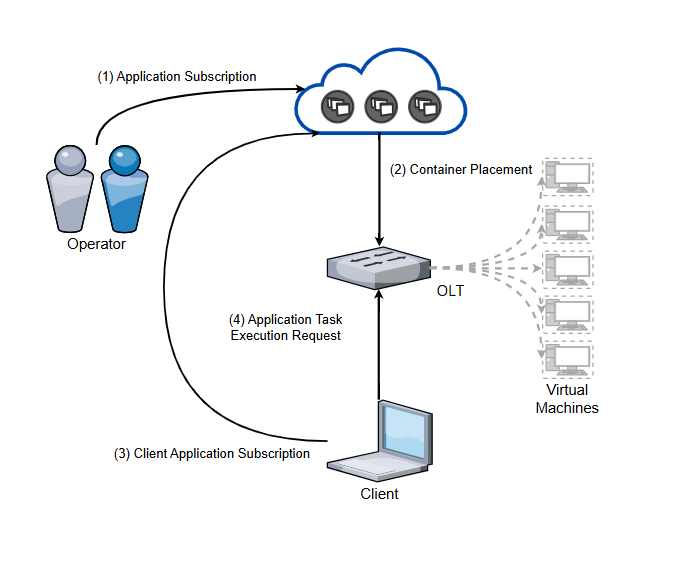}
  \caption{Service subscription model in \toolname{}. Operators deploy applications as containers at the Cloud level, while clients subscribe to services and issue tasks through the Edge.}
  \label{fig:sub_model}
\end{figure}

In this paradigm, operators deploy applications as long-running services, while clients subscribe to them and subsequently issue tasks (Figure~\ref{fig:sub_model}).
The \toolname{} event model reflects this separation: the \texttt{Simulation Manager} schedules container lifecycle events independently of task arrivals, and the \texttt{Task Orchestration} module maintains two distinct flows, one for managing service placement and another for computing per-task offloading decisions.

Decoupling service placement from task offloading requires introducing dedicated abstractions, different from previous simulators. 
In \toolname{}, containers are now first-class entities, distinct from tasks. They are instantiated through a dedicated \texttt{ContainerGenerator} that supports both \textit{shared} instances (i.e., container instances can serve multiple clients) and \textit{private} instances (i.e., each instance serves a single client). Applications explicitly reference container attributes in addition to task parameters, while a new \texttt{ContainerTransferProgress} class captures the impact of container deployment on network bandwidth. Core modules were adapted to reflect this separation. The \texttt{Task Generator} now produces tasks independently of containers, with private tasks only enabled after the associated container is placed. The \texttt{Simulation Manager} initializes container placement prior to task generation, reflecting operator-driven deployment followed by client-driven execution. The \texttt{Server Manager} maintains per-node container queues, constrained by memory and storage resources. Orchestration responsibilities are split between a cloud-level \texttt{ContainerOrchestator}, which manages placement and lifecycle events, and an edge-level \texttt{TaskOrchestrator}, which handles runtime offloading in line with current container distribution.

\begin{figure}[t]
    \centering
    \includegraphics[width=0.8\linewidth]{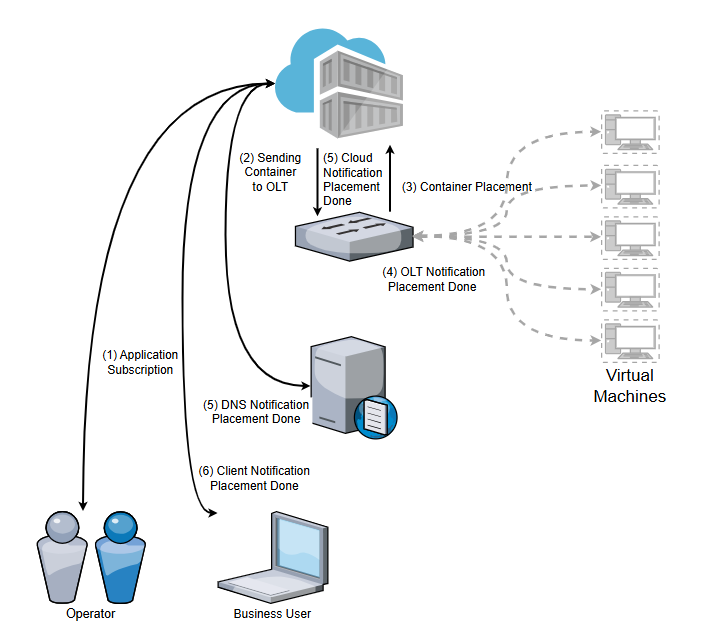}
    \caption{Routing of service placement requests in \toolname{}. Operator deployments are sent to the Cloud orchestrator, which selects target nodes (OLTs/ONTs) and distributes container images accordingly. Placement updates are propagated to DNS brokers and subscribed clients.}
    \label{fig:placement_routing}
\end{figure}

\begin{figure}[t]
    \centering
    \includegraphics[width=0.8\linewidth]{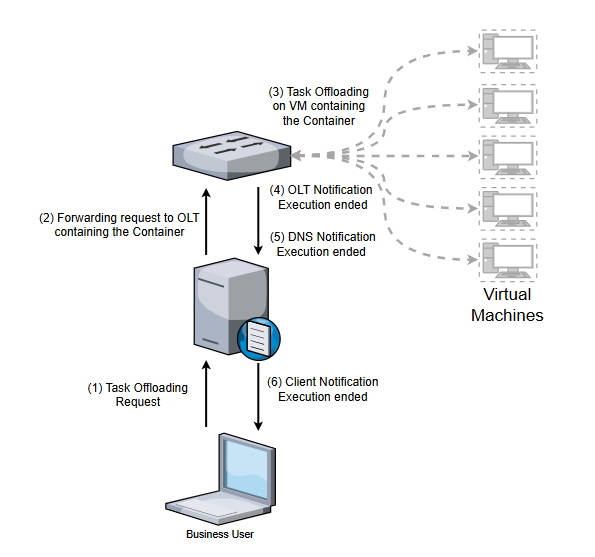}
    \caption{Routing of task execution requests in \toolname{}. Clients submit tasks to the nearest OLT-side DNS broker, which selects the optimal service replica based on placement state, resource availability, and offloading policy.}
    \label{fig:execution_routing}
\end{figure}

\toolname{} adopts a request routing model designed to reflect hierarchical orchestration. The simulator models two types of requests: (1) requests for service placement, made by service operators, in which container images flow from the Cloud to OLTs, with notifications propagated to DNS brokers and clients (Figure~\ref{fig:placement_routing}); (2) requests for task execution, in which end-users send requests to a service, using DNS brokers to resolve the container instance for processing the request (Figure~\ref{fig:execution_routing}).

By introducing hierarchical orchestration, \toolname{} captures the separation of placement and execution required by GENIO, and enables fine-grained analysis of orchestration policies under realistic container lifecycles and load dynamics.

\subsection{Orchestration Strategies: Placement}
\label{sec:placement}
The following four service placement algorithms are implemented in \toolname{}. Each algorithm defines a heuristic for selecting the best \textit{computing node}, where a container should be deployed.

\paragraph{Round Robin}
The algorithm cycles through the list of available computing nodes and assigns each new container to the next node in order, without considering resource capacity, load, or network conditions. All nodes are treated uniformly. This approach is computationally lightweight and deterministic; however, it can still overload nodes with less computational capacity and create suboptimal paths between users and containers. Round Robin scheduling is widely used in operating systems and cloud orchestration systems~\cite{tanenbaum}.

\paragraph{CPU-Greedy}
It applies a per-core, count-based greedy rule that selects the node with the lowest ratio between the number of currently assigned containers $C_i$ and the total number of CPU cores $N_i$. In the event of a tie, the algorithm prefers the node with a higher number of cores. This heuristic performs fast, low-overhead decisions that roughly balance container counts in proportion to core counts. However, the objective is not resource- or utilization-aware, since it disregards available memory, storage, and per-core MIPS, potentially leading to suboptimal placements under heterogeneous nodes and varying container demands. Formally, the selected node minimizes the ratio defined in Equation~\ref{eq:cpu_greedy}. This heuristic is first introduced in this work.

\begin{equation}
F_i = \frac{C_i}{N_i}
\label{eq:cpu_greedy}
\end{equation}

\paragraph{Trade-Off}
It selects the node that minimizes a topology-weighted queueing cost function that approximates the expected completion time. For each node $i$, the cost $F_i$ grows linearly with the current number of assigned containers $C_i$ and the size of the incoming task $S_t$, and decreases with the node's per-core processing power $MIPS_i$ (Million Instructions per Second). A topology-dependent weight $t_i$ penalizes nodes in the cloud layer and favours nodes in the edge and far-edge layers (i.e., ONTs and OLTs) in the selection process. This heuristic is resource-aware (in terms of per-core MIPS and task size) and topology-aware. Still, it overlooks memory and storage availability as well as total core count, which may lead to under-utilization of multi-core nodes. The node with the lowest cost is selected for allocation. Formally, the cost function is defined in Equation~\ref{eq:trade_off}. This heuristic was originally introduced in the PureEdgeSim simulator~\cite{mechalikh2019pureedgesim}. 

\begin{equation}
F_i = (2C_i + 1) \cdot t_i \cdot \frac{S_t}{\text{MIPS}_{i}}
\label{eq:trade_off}
\end{equation}

\paragraph{Multi-Objective}
It maximizes a resource-aware and topology-aware availability score over feasible nodes. For each node $i$, the score $S_i$ aggregates normalized available resources $R$, including RAM, storage, number of CPU cores, and per-core MIPS, and penalizes nodes with a large number of active containers $C_i$. A topology bias term $t_i$ assigns a lower penalty to nodes in the edge and far-edge layers (e.g., ONTs and OLTs) and a higher one to nodes in the cloud layer. This weighted-sum approach captures both computational heterogeneity and topology preference, albeit at the cost of higher decision overhead and complexity compared to the CPU-greedy approach. The node maximizing the score defined in Equation~\ref{eq:multi_objective} is selected for allocation. This heuristic is first introduced in this work.

\begin{equation}
S_i = 
\sum_{R \in 
    \substack{
        \{\text{RAM},\,\text{storage},\\
        \text{cores},\,\text{MIPS/core}\}
    }
}
\left(w_R\,\frac{R_{i,\text{avail}}}{R_{i,\text{tot}}}\right)
- C_i - t_i.
\label{eq:multi_objective}
\end{equation}

Each of these algorithms has three \textit{variants}, which differ in how they determine the \textit{scope} of the placement decision. 

\bigbreak

\begin{itemize}
    \item \textit{Standard variant} applies the algorithm globally across all computing nodes in the topology (including those in the cloud, OLTs, and ONTs) to directly select the best node for placement. The other two variants (\textit{Latency-based} and \textit{Rate-based}) adopt a two-level decision process that involves multiple OLTs and ONTs. First, they evaluate all potential hosts, that is, all OLTs and ONTs in the simulated topology, to determine which ones are the most suitable to host the container. 

    \item \textit{Latency-based variant} analyzes, for each OLT, the mean network latency between containers currently deployed on that OLT and end-user devices associated with that container, selects the OLTs with the lowest average latency, and among them chooses the one currently hosting the fewest containers of the same type. 

    \item \textit{Rate-based variant} measures, for each OLT, how many edge devices are already connected and how many container copies of the same type are already deployed, then selects the OLT with the lowest ratio between deployed copies and associated devices.
\end{itemize}

\bigbreak
After this host-selection phase, the chosen OLT (or ONT) becomes the local placement scope. The corresponding base algorithm (Round Robin, Trade-Off, CPU-Greedy, or Multi-Objective) is applied within that host’s associated computing nodes (e.g., VMs in the case of OLTs or the physical node itself in the case of ONTs) to select the final target node for container deployment. In summary, the Standard variant optimizes placement globally across the entire infrastructure. In contrast, the Latency-based and Rate-based variants first identify the best host among multiple OLTs and ONTs and then optimize placement locally within that host according to network proximity and load proportionality, respectively.

\subsection{Orchestration Strategies: Offloading}
\label{sec:offloading}
The following task offloading algorithms are implemented in \toolname{}. Each algorithm defines how user requests (tasks) are routed to containers that run the required service.

\paragraph{Round Robin}
This algorithm distributes incoming tasks across all container instances of the same application in cyclic order. Each new task is forwarded to the next container in the sequence, independently of the container’s location, load, or execution delay. This ensures fairness and negligible scheduling overhead, but does not exploit differences in network latency or node performance, which may result in suboptimal offloading decisions.

\paragraph{Best Latency}
This algorithm assigns each task to the container instance that offers the lowest end-to-end network latency from the task’s originating edge device. If multiple containers have the same minimum latency, it selects among them the one with the fewest currently assigned tasks. This strategy is network-aware but largely load-agnostic, balancing load only when latency ties occur. This heuristic is first introduced in this work.

\paragraph{Best Delay}
This algorithm extends the best latency strategy by accounting for both communication delay and an approximation of execution delay on the destination node. For each container instance $i$ implementing the required application, it computes the network latency $L_i^{net}$ from the task's originating edge device to the container's hosting node. It estimates the execution delay by summing the compute demand $S_t$ of all tasks already queued on that node (plus incoming tasks), dividing by the node's per-core processing capacity $MIPS_i$, and normalizing by the number of CPU cores $N_i$. This strategy is explicitly delay-aware, combining network and processing delays. The task is dispatched to the container that minimizes the predicted delay, as defined in Equation~\ref{eq:best_delay}. This heuristic is first introduced in this work.

\begin{equation}
D_i = L_i^{net} + \frac{\sum_{t \in Q_i} 
(\frac{S_t}{MIPS_i}) + \frac{S_{new}}{MIPS_i}}{N_i}
\label{eq:best_delay}
\end{equation}

Each offloading algorithm is evaluated in two \textit{modes}, which determine how often the offloading destination is recomputed.

\begin{itemize}
    \item \textit{Static mode} associates each edge device with a fixed offloading node the first time it generates a task. This mapping is stored and reused for all subsequent tasks from that device. As a result, offloading decisions incur negligible runtime overhead but cannot adapt to changes in network latency, node load, or task queue lengths. Static mode, therefore, prioritizes stability and speed over responsiveness. 

    \item \textit{Dynamic mode} recomputes the offloading decision for every incoming task. The system evaluates the selected offloading algorithm (Round Robin, Best Latency, or Best Delay) each time by inspecting current network conditions, container placement, and queue state. This enables fine-grained responsiveness to transient load and latency fluctuations, but introduces additional control-plane computation. 
\end{itemize}

\subsection{Workload and Users Modeling}
In previous simulators, such as PureEdgeSim, the workload is represented by tasks, which are modeled using parameters like task size (e.g., in terms of CPU instruction counts needed to process the task) and rate of task arrival. However, as previously discussed, this approach does not fit well with long-running services, which process multiple requests once they are instantiated, and which are deployed in multi-tenant infrastructures. Moreover, real-world industrial applications often involve various users with heterogeneous behavioral profiles.

\toolname{} provides a richer modeling of workloads and users, by clearly separating services from their users, and allowing multiple user behavioral profiles. In particular, this modeling involves the \texttt{Task Generator} and the \texttt{Scenario Manager} modules. On the one hand, application services are modeled as containers, with configurable properties such as memory and storage requirements, bandwidth footprint, number of replicas, and whether a container instance is shared among multiple users. Containers persist as long-lived services, enabling evaluation of lifecycle events such as creation, scaling, and removal. 

Users, on the other hand, are defined through separate profiles, which specify application type, access pattern (random, bursty, or periodic), request rates, and temporal behavior (start time, active periods, idle intervals).  This fine-grained configuration enables the reproduction of realistic workload dynamics, including diurnal cycles, sudden bursts of demand, and session-oriented behavior. By decoupling user activity from application instances, \toolname{} enables heterogeneous, multi-tenant environments where users compete for shared services.

This design supports simulations of stressful and highly variable workloads, such as bursty video on-demand streaming and ML-based event processing patterns. Together, these extensions provide a much richer workload model than PureEdgeSim, enabling practitioners to investigate orchestration strategies under conditions that mirror real-world variability and peak loads.

\bigbreak
\Definition{Overall, the enhancements provided by \toolname{} bridge the gap between edge computing simulators and the unique requirements of PON-enabled infrastructures. This approach enables designers and researchers to evaluate orchestration strategies under realistic network, virtualization, and application conditions.}

\section{Experimental Analysis}
\label{sec:exp}
This section presents experiments using \toolname{} to evaluate how well PON-enabled, hierarchical infrastructures can support edge computing workloads. First, we experiment on \textit{capacity planning}, where we use \toolname{} to determine the minimum computational resources required to sustain full task completion under application-specific latency constraints. Second, we conduct an experiment to \textit{compare orchestration policies}, where we use \toolname{} to assess how different placement and offloading strategies affect overall system performance.

These experiments represent two relevant use cases for the proposed tool. The experiments use \toolname{} to reproduce PON-based edge computing deployments, where we measure both the \emph{Task Success Rate (TSR)}, i.e., the fraction of tasks completed within the application-specific latency bound, and the \emph{end-to-end latency}, defined as the time between the submission of a task from the user and its completion.

\subsection{Application Scenarios}
\label{sec:app_scenarios}
Different edge applications are modeled by tuning the simulation parameters, which are derived from prior works that emulate realistic edge computing scenarios~\cite{cesarano2025genio, mechalikh2019pureedgesim, enhancing2024, collab}. These parameters are confirmed by empirical data provided by the industrial partners of the GENIO project. We model several application scenarios with distinct computational intensities and latency service-level objectives (SLOs). These scenarios capture diverse workload characteristics, with some compute-demanding tasks and others involving data-intensive exchanges. In contrast, others generate frequent request bursts, thereby reflecting the heterogeneity typical of real-world edge deployments.

For the experiment on capacity planning, we consider five different applications: 

\begin{enumerate}
    \item[S1] \textit{Smart City}: 128 users representing connected streetlight nodes supporting urban monitoring services;
    \item[S2] \textit{E-Health}: 10 users representing remote patients equipped with continuous health monitoring devices;
    \item[S3] \textit{Smart Building}: 20 users representing surveillance and analytics endpoints within an intelligent facility; 
    \item[S4] \textit{Sports Streaming}: 60 users representing simultaneous viewers of live multimedia content; 
    \item[S5] \textit{Video Gaming}: 80 users representing participants in interactive cloud-gaming sessions.
\end{enumerate}

In addition, for orchestration strategy evaluation, we define a \textit{Mixed scenario} that combines all five baseline applications into a single multi-tenant setup with a total of 298 users, generating heterogeneous and time-varying demand representative of realistic industrial deployments. Table~\ref{tab:config} summarizes the main parameters of each scenario, including task rates, latency SLOs (column “Max latency”), computational demand (columns “Request size” and “Task length”), response size, and number of active users, thus defining the five baseline workloads used in our experiments.

\begin{table}[h]
\caption{Simulation parameters for the application scenarios}
\label{tab:config}
\centering
\small
\begin{tabular}{lccccc}
  \toprule
   & \textbf{S1} & \textbf{S2} & \textbf{S3} & \textbf{S4} & \textbf{S5} \\
  \midrule
  Active Users & 128 & 10 & 20 & 60 & 80 \\
  Task rate (task/min) & 2 & 60 & 60 & 20 & 180 \\
  Max latency (s) & 0.5 & 0.05 & 0.2 & 0.5 & 0.05 \\
  Task length (MI) & 500 & 1,000 & 5,000 & 5,000 & 100 \\
  Request size (KB) & 1 & 10 & 750 & 750 & 10 \\
  Response size (KB) & 10 & 10 & 500 & 500 & 10 \\
  \bottomrule
\end{tabular}%
\end{table}

\subsection{Capacity planning}

\paragraph{Objective}
We aim to determine the minimum computational capacity required to sustain 100\% TSR for each application when deployed on a single OLT in isolation, and to quantify how extending computation to the far edge (ONTs) affects both TSR and latency. The success criterion is the smallest OLT capacity that maintains 100\% TSR throughout the experiment duration while meeting the application’s latency SLO.

\paragraph{Experimental Setup}
We simulate a single-OLT topology centered in the simulation region, with its associated ONTs representing the far-edge layer. The OLT models a physical x86-class host with 8 cores, serving as the upper-tier edge node, consistent with the hardware configuration already deployed in OLT systems by our industrial partner~\cite{selta_volt}. Each core's computational capacity, expressed in MIPS, is varied across experiments to emulate different commercial CPU classes (see the next paragraph on Methodology). On top of this physical OLT, we instantiate four VMs, each allocated two of the physical cores. The aggregate computational capacity of the OLT is therefore divided among these VMs, whose per-core MIPS values are adjusted to reflect the characteristics of the CPU being emulated. Each ONT emulates a Raspberry Pi 4 Model B–class device with four cores at 12{,}000~MIPS/core, representing far-edge resources close to end users. The number of instantiated ONTs matches the active user count in the selected application scenario (Section~\ref{sec:app_scenarios}).

We evaluate two deployment configurations: (1) \textit{Edge-only}, where all tasks are executed on OLT VMs; and (2) \textit{Far-Edge + Edge}, which allows ONTs to execute tasks locally when resources are sufficient.

All experiments use single-application workloads to isolate capacity effects and avoid inter-application interference. Container placement follows the Trade-Off policy, and task offloading uses a Round-Robin strategy, providing a conservative baseline for capacity estimation. Each simulation runs for 300 minutes of simulated time.

\paragraph{Methodology} 
To evaluate the OLT's computational capacity requirements, we systematically vary its total capacity processing power to match the performance of five representative commercial x86 CPU classes. For each class, we assign a per-core MIPS value derived from the Sandra Dhrystone benchmark~\cite{sandra_drystone} and scale it across the OLT's eight cores. The ONT resources remain constant throughout.

For each application and deployment model, we proceed as follows. We configure the OLT with the target CPU capacity and \toolname{} instantiate resources to simulate (including VMs); initialize container placement and execute the workload; record TSR and mean end-to-end latency during the simulated time, over five independent runs; average results across runs to obtain stable statistics.

\begin{figure}[!t]
    \centering

    \includegraphics[width=0.5\textwidth,trim=2cm 0 0 0,clip]{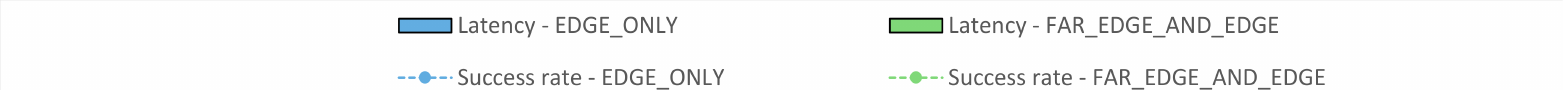}\par\medskip
    \includegraphics[width=0.48\textwidth]{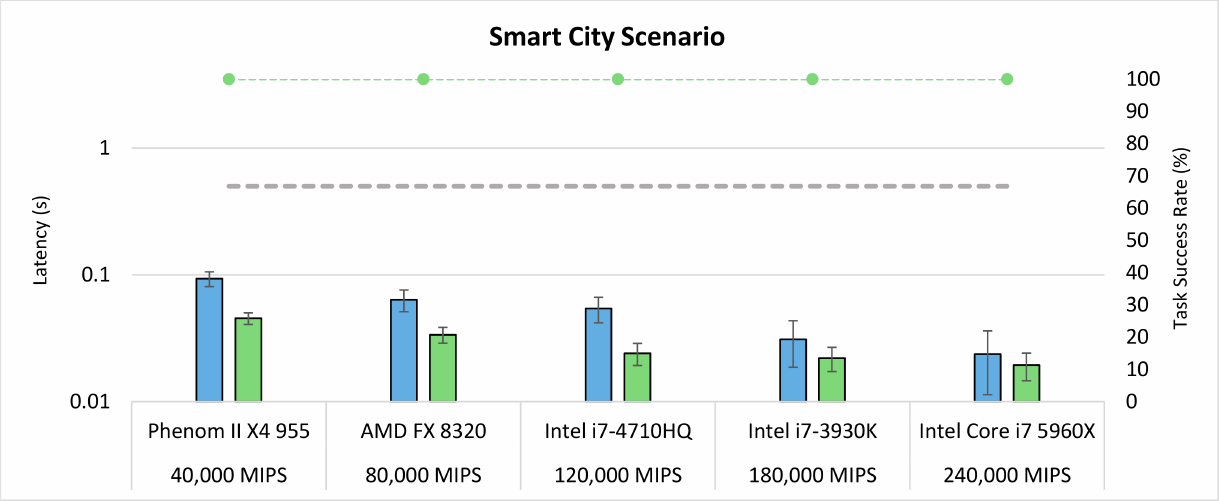}\par\medskip
    \includegraphics[width=0.48\textwidth]{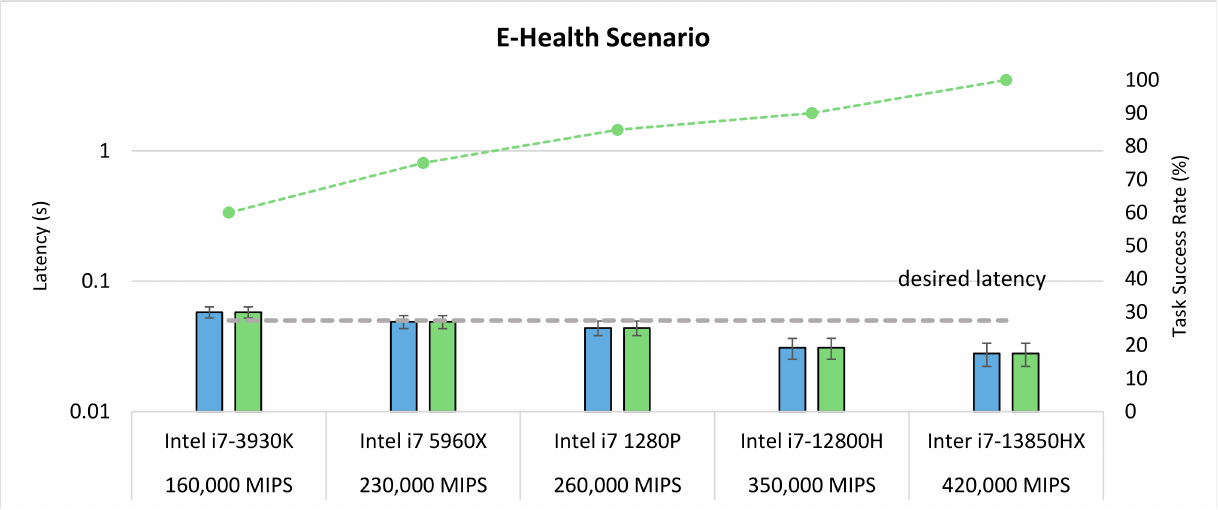}\par\medskip
    \includegraphics[width=0.48\textwidth]{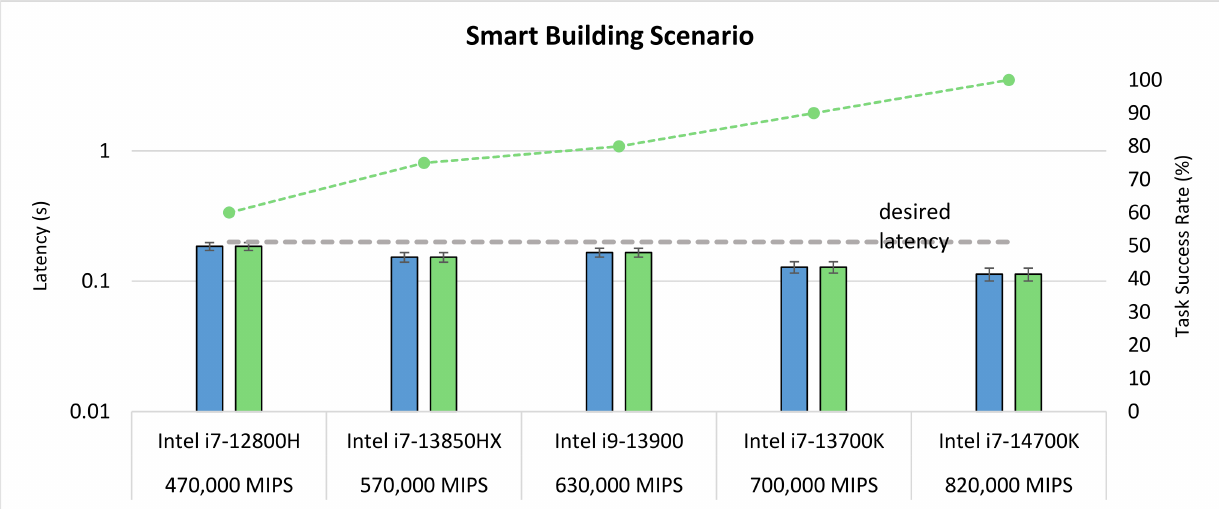}\par\medskip
    \includegraphics[width=0.48\textwidth]{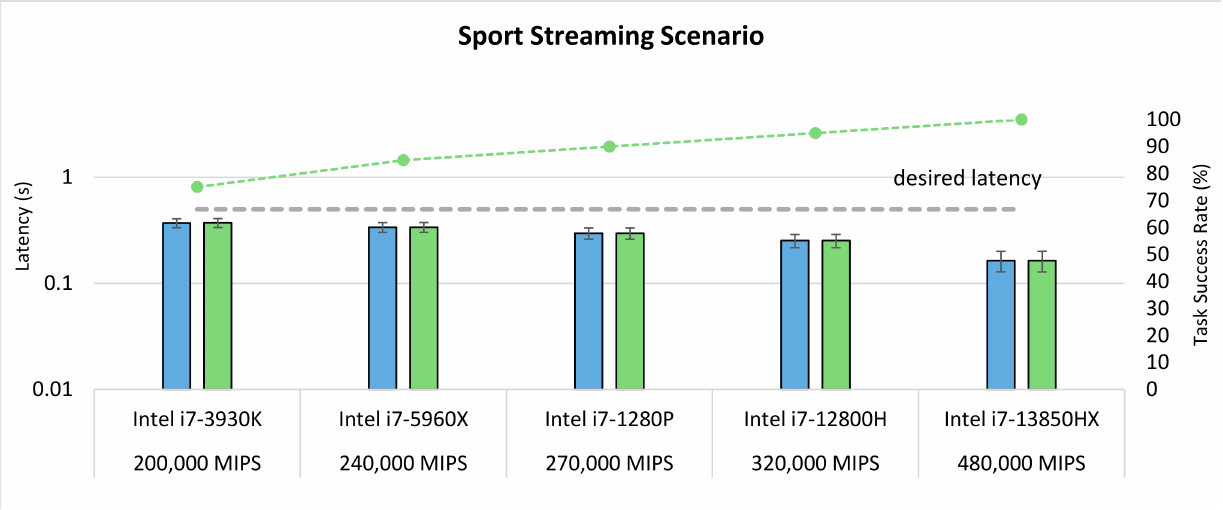}\par\medskip
    \includegraphics[width=0.48\textwidth]{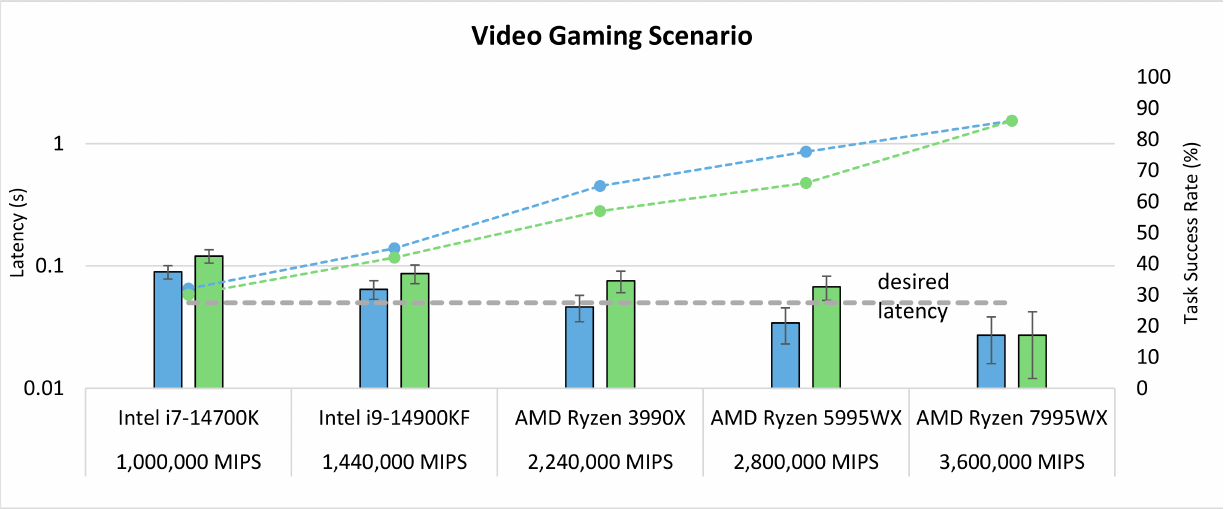}

    \caption{TSR and end-to-end latency across CPU capacity sweeps for Edge-only and Far-Edge + Edge deployments, evaluated over five application scenarios. When only one TSR curve is shown, the two deployments produce overlapping results.}
    \label{fig:simulation_results}
\end{figure}

\paragraph{Results}
Figure~\ref{fig:simulation_results} presents, for each application scenario:
(1) mean latency achieved by the different CPU classes for both Edge-only and Far Edge + Edge deployments (bars), and 
(2) task success rate (TSR) achieved under the same configurations (overlaid as lines).
These values are reported for different CPU classes. The computation capacity of each CPU is provided on the bottom of each plot.

Since applications differ in computational load and latency sensitivity, the resource capacity needed to meet SLOs varies across scenarios. 
Overall, lighter workloads benefit from far-edge execution, while compute-intensive or latency-sensitive applications remain OLT-bound. 
Smart City always achieves 100\% TSR due to its low computational demand (500 MI per task). Allowing ONTs to execute tasks further reduces latency by supplementing OLT capacity and minimizing network traffic. 
E-Health, Smart Building, and Sport Streaming exhibit monotonic improvement as the capacity of the OLT increases, as TSR increases, and latency decreases. Transitioning from Edge-only to Far-Edge + Edge provides marginal gains, since these applications typically execute at the OLT to meet moderate SLOs (0.05–0.5 s) with stable compute performance.
Video Gaming poses the most demanding scenario. At low OLT capacities, Far-Edge + Edge increases task failures, as ONTs admit tasks but are frequently unable to process all of them, resulting in inflated service times and missed deadlines. In contrast, Edge-only avoids overloads and achieves lower latency under the same OLT capacities.

\Definition{\textbf{Takeaways}: (i) Capacity planning is application-specific; simulations can guide the selection of the minimal hardware resources needed to meet application requirements. (ii) Distributing the load on the far-edge is effective to reduce latency for light workloads with sufficient ONT capacity. Still, more complex workloads with strict time constraints do not necessarily benefit from the far edge.
(iii) For heavy computational workloads (e.g., gaming), introducing the far edge can expose to task failures (i.e., TSR degradation), which needs to be enforced through conservative task admission.}

\subsection{Comparison of Orchestration Policies}

\paragraph{Objective}
We quantify the impact of container placement and task offloading strategies on TSR and latency under heterogeneous, bursty workloads in a multi-OLT deployment. The goal is to identify policy combinations that sustain high TSR while minimizing latency.

\paragraph{Experimental Setup}
We simulate a topology with three OLTs, and devices on the far edge evenly distributed among the OLTs. Each OLT provides 14 physical cores (95{,}000~MIPS/core) for a total of 1.33~million~MIPS per OLT. The far-edge layer consists of ONTs following the same hardware modeling as described in the previous experiment. In this case, each OLT hosts 7 VMs (two cores each). The far-edge layer includes one ONT per user, for a total of 298 ONTs corresponding to all users in the Mixed workload (Section~\ref{sec:app_scenarios}).

We evaluate the same two deployment models (Edge-only and Far-Edge+Edge) used previously. The workload, however, is now heterogeneous, combining all five application scenarios to generate concurrent and bursty task arrivals representative of realistic multi-tenant conditions.

\paragraph{Methodology}
For each simulation, a placement algorithm and an offloading algorithm are selected together with their respective variants. We evaluate 12~placement strategies (4~algorithms~$\times$~\{standard, latency-based, rate-based\}) and 6~offloading strategies (3~algorithms~$\times$~\{static, dynamic\}), totaling 72~policy combinations. Each combination is executed five times per deployment model, yielding 720~simulations in total. Placement and offloading mechanisms are implemented as described in the Sections~\ref{sec:placement} and ~\ref{sec:offloading}.

For each run, we record the TSR and mean latency. TSR is computed as the ratio of completed tasks to the total number of submitted tasks, providing a global measure of orchestration efficiency. Latency is collected per application and normalized by its respective target SLA (Table~\ref{tab:config}) to ensure comparability across applications with different latency scales. The normalized values are then averaged to yield a single representative latency per policy combination ($L_{\text{norm}}$). Consequently, latency values reported here represent SLA-normalized averages and are not directly comparable to the raw latency results from the provisioning experiment.

\begin{figure*}[t]
  \centering

  \includegraphics[width=\linewidth,trim=0cm 1.5cm 0cm 1.5cm,clip]{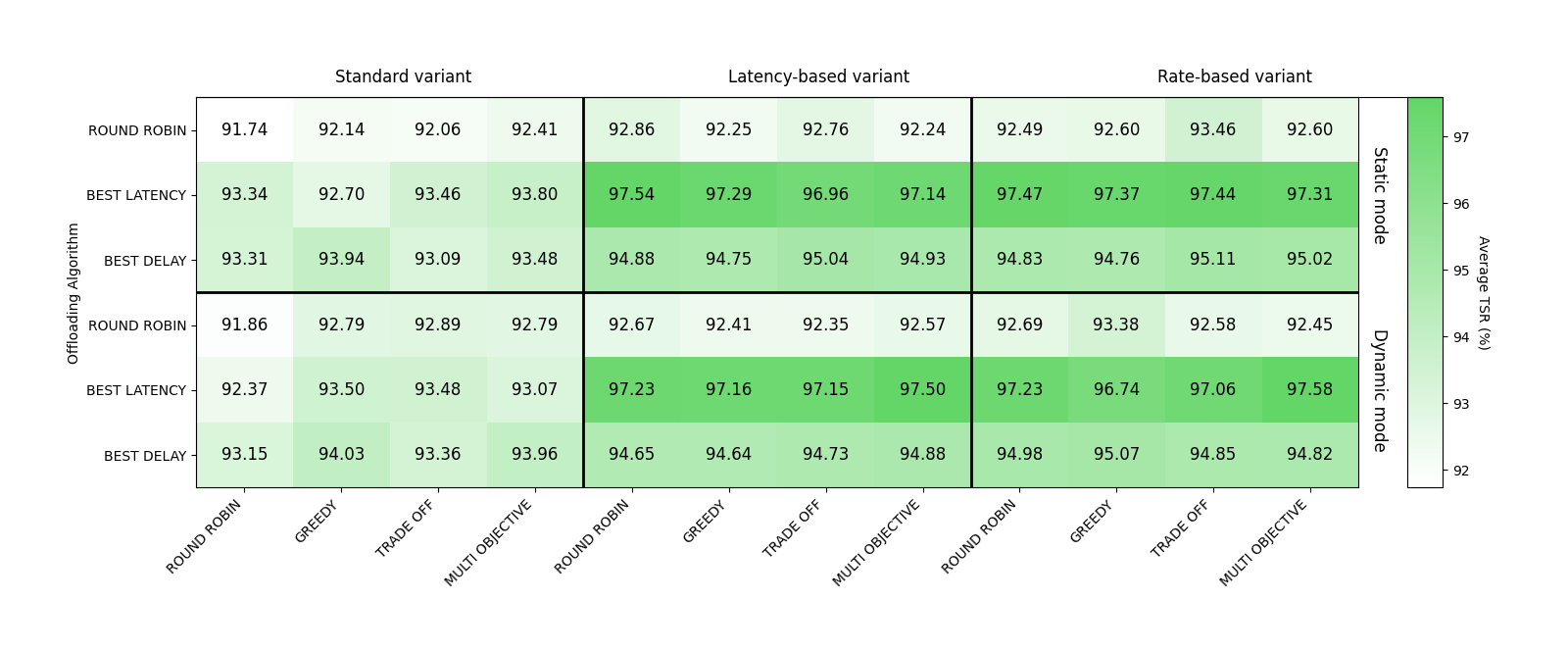}\\
  \includegraphics[width=\linewidth,trim=0cm 1cm 0cm 1.5cm,clip]{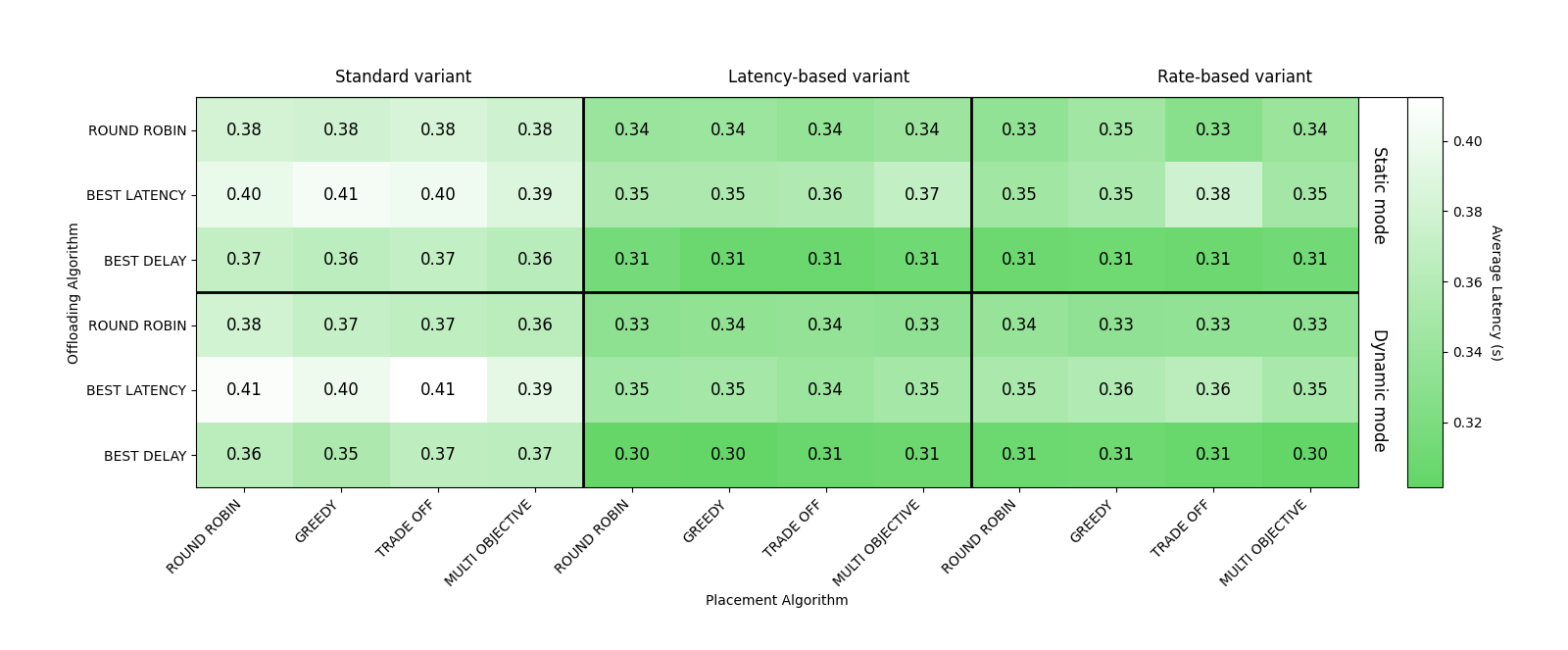}

  \caption{Comparison of TSR (top) and normalized latency (bottom) across all combinations of offloading and placement strategies when used in \textit{edge-only} deployment. Each cell reports the mean over five runs.}
  \label{fig:edge_only_exp1}
\end{figure*}

\paragraph{Results for Edge-Only}
Figure~\ref{fig:edge_only_exp1} presents heatmaps of TSR and normalized latency for all combinations of offloading and placement strategies obtained under the \textit{edge-only} deployment; rows list the offloading algorithms (Round Robin, Best Latency, Best Delay; static/dynamic modes) and columns group the placement algorithms (Round Robin, Greedy, Trade Off, Multi-Objective; Standard/Latency-based/Rate-based variants). Each cell reports the mean value over five runs. All policies under the Edge-only configuration maintain full SLA compliance, with normalized latencies ranging between $0.30$ and $0.41$ (mean $\approx0.35$).
The narrow latency range indicates that, in a compute-rich environment, placement and offloading choices have only a secondary impact on performance.
Among offloading variants, \textit{Best Delay (BD)} consistently achieves the lowest normalized latencies (around $0.30$--$0.31$), owing to its explicit estimation of end-to-end service time (\(L_{net}+Q_{wait}+S_t/\text{MIPS}\)).
\textit{Best Latency (BL)} also performs well but yields slightly higher delay ($\sim0.34$--$0.36$) since it ignores queuing effects and may offload tasks to temporarily busy nodes.
For placement, both \textit{latency-based} and \textit{rate-based} variants reduce the average latency by roughly 7--10\% compared to their \textit{standard} counterparts by biasing container deployment toward nearby or lightly loaded nodes. 
Dynamic offloading yields a modest additional benefit ($\Delta L_{\text{norm}}\approx-0.01$), as per-task adaptation smooths transient load peaks across VMs.

Regarding the success of tasks, all Edge-only configurations achieve high TSR between $91.7\%$ and $97.6\%$ (mean $\approx94.4\%$).
Differences in TSR follow offloading strategy more than placement: \textit{BL} variants deliver the highest completion rates (up to $97.6\%$), while \textit{BD} and \textit{RR} are slightly lower ($\sim94$--$95\%$).
This reflects BL’s bias toward low-latency OLT nodes, which minimizes queueing losses under the high workload intensity of this scenario.
Dynamic offloading slightly improves TSR but with marginal average gains ($<1$ percentage point).
Latency- and rate-based placement variants raise TSR by about 1--2\% relative to standard ones through improved task-to-container locality.
Overall, the Edge-only configuration confirms that with sufficient OLT capacity, all orchestration strategies can satisfy SLA constraints, and performance differences mainly stem from the adaptability of the offloading policy rather than placement limitations.

\begin{figure*}[h]
  \centering
  \includegraphics[width=\linewidth,trim=0cm 1.5cm 0cm 1.5cm,clip]{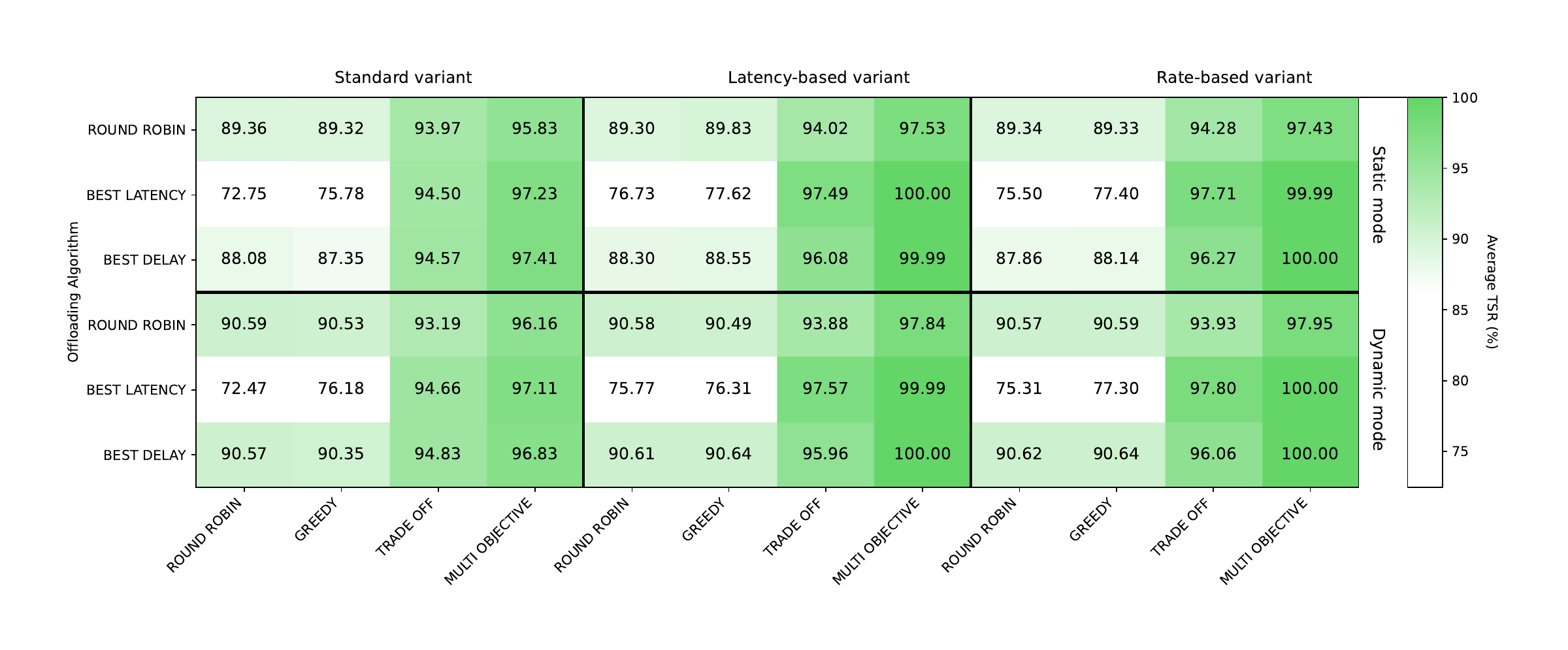}\\
  \includegraphics[width=\linewidth,trim=0cm 1cm 0cm 1.5cm,clip]{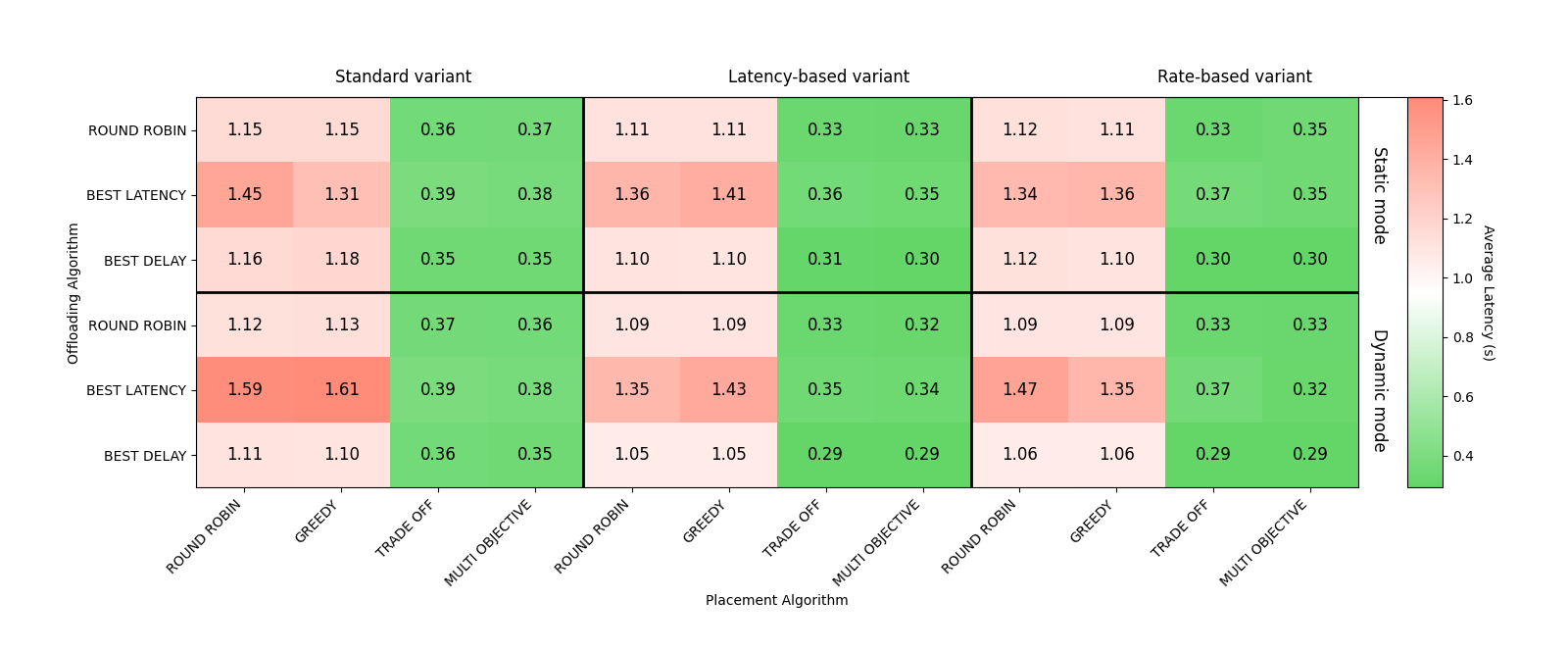}

  \caption{Comparison of TSR (top) and normalized latency (bottom) across all combinations of offloading and placement strategies when used in \textit{Far-Edge + Edge} deployment. Each cell reports the mean over five runs.}
  \label{fig:edge_faredge_exp1}
\end{figure*}

\paragraph{Results for Far-Edge+Edge}
Figure~\ref{fig:edge_faredge_exp1} shows the corresponding heatmaps for the Far-Edge + Edge deployment.
Introducing the Far-Edge layer heavily changes latency dynamics. 
Normalized latencies now span a wider range ($0.29$--$1.59$), and only about half of the policy combinations satisfy the SLA condition (\(L_{\text{norm}}<1\)).
Average latency across all algorithms increases to approximately $0.78$, exhibiting a bimodal distribution: \textit{Multi-Objective} and \textit{Trade-Off} placements consistently remain below SLA (mean $L_{\text{norm}}\approx0.34$), while \textit{Round-Robin} and \textit{CPU-Greedy} variants incur persistent violations ($L_{\text{norm}}\approx1.1$–$1.6$).
This divergence stems from how each placement algorithm leverages heterogeneous compute resources.
Latency- and rate-based variants clearly outperform the base algorithms, reducing latency by roughly 8–10\% on average, by aligning container placement with node proximity and available capacity.
In contrast, the base algorithms often distribute containers arbitrarily, causing tasks to traverse multiple network hops or accumulate at ONTs, which have lower computational capacity. 
Offloading decisions amplify these effects: static policies perform consistently worse than dynamic ones, as their fixed routing cannot respond to transient congestion and load imbalances. Dynamic offloading lowers mean latency by 5–10\% across most strategies, especially when paired with delay-aware offloading.

TSR trends mirror the latency distribution. 
The mean TSR across all policies decreases to about $91\%$, but again splits by algorithm: \textit{Multi-Objective} and \textit{Trade-Off} placements achieve near-perfect reliability (up to $100\%$ TSR), while \textit{Round-Robin} and \textit{Greedy} collapse below $80\%$.
Policies combining latency- or rate-based placement with \textit{dynamic BD} offloading dominate the high-performance region.
In contrast, proximity-only (\textit{BL}) strategies show the lowest TSR, down to $72\%$, as they over-admit tasks to resource-constrained ONTs, resulting in service-time overruns and deadline misses.
These findings confirm that, in heterogeneous edge hierarchies, compute-awareness outweighs mere proximity: effective orchestration must pair placement guided by node capacity with offloading that reacts to current service delay.

\paragraph{Comparison between deployment models}
Comparing the two deployment models shows that adding far-edge nodes does not necessarily improve latency and task success, unless appropriate orchestration policies are applied to account explicitly for the Far-Edge+Edge deployment model. 
The average variation in terms of success rate ($\Delta$TSR) from Edge-only to Far-Edge+Edge is $-3.3$ percentage points, and the average $\Delta L_{\text{norm}}$ is $+0.43$. 
However, the best policies, which are the \textit{Multi-Objective} or \textit{Trade-Off} placement with dynamic \textit{Best Delay} offloading, reverse this trend, achieving $\Delta$TSR $= +5$ points and $\Delta L_{\text{norm}} = -0.06$. 
These configurations succeed because they exploit ONTs only when the predicted service delay, including processing cost, remains below the OLT alternative. 
Conversely, naïve policies (e.g., \textit{Round-Robin} placement with static \textit{Best Latency} offloading) significantly degrade with respect to both metrics (\(\Delta\)TSR $-20$ points, \(\Delta L_{\text{norm}} +1.0\)). 
Overall, the results demonstrate that far-edge computation improves performance only under adaptive, delay-aware orchestration; otherwise, proximity to the user paradoxically increases end-to-end latency due to queuing effects.

The observations above can be summarized through the following key insights regarding orchestration policies in PON-enabled edge infrastructures. In particular, the experiments show that the best performance is achieved by combining compute-aware placement strategies such as \textit{Multi-objective} or \textit{Trade-Off} with \textit{dynamic Best Delay} offloading, which jointly account for node capacity, queueing delay, and network latency.

\Definition{\textbf{Takeaways}: (i) Introducing the Far-Edge layer amplifies the performance gap between compute-aware and proximity-based policies: only strategies that jointly account for node capacity and network delay sustain SLA compliance.
(ii) Multi-objective and trade-off placements, when combined with dynamic delay-aware offloading, consistently achieve high TSR and low latency across heterogeneous conditions.
(iii) In contrast, proximity-driven or static policies (e.g., BL with Round-Robin placement) overload ONTs and violate SLA guarantees, confirming that in heterogeneous infrastructures, compute-awareness outweighs mere proximity.}

\section{Simulator Scalability and Performance}
\label{sec:scalability}

\paragraph{Objective}
\toolname{} targets scenarios with multiple OLTs, hundreds of ONTs, and bursty multi-tenant worklods. To assess the practical scalability of \toolname{}, we evaluate how the simulator's runtime and memory usage evolve as the size of the simulated infrastructure and number of users increase. 

\paragraph{Experimental Setup}
All scalability experiments use the \textit{Mixed scenario} described in Section~\ref{sec:app_scenarios}, which combines all application workloads to generate heterogeneous and bursty traffic patterns. Two scaling dimensions are explored.

In the first experiment, the number of users is fixed at 1{,}000, with each user represented by one ONT; while the number of OLTs varies from 10 to 100 in increments of 10. This configuration evaluates how the simulator scales with the size of the access network infrastructure.

In the second experiment, the number of OLTs is fixed at 100 while the number of users varies from 100 to 1{,}000 in increments of 100. This configuration evaluates how the simulator scales with the number of edge devices and workload generators.
For each simulation run, we measure (i) the total simulation runtime and (ii) the maximum resident set size (RSS) of the simulator process, representing peak memory consumption. Each configuration is executed three times, and the reported values correspond to the average across runs.

All simulations were executed on a workstation equipped with an Intel Core i7 processor (8 cores, 3.2\,GHz), 32\,GB of RAM, running Ubuntu 22.04.

\paragraph{Results}
Table~\ref{tab:scalability_olts} reports the results when varying the number of OLTs while keeping 1{,}000 users fixed. The simulator runtime increases moderately from 142~s to 155~s as the number of OLTs grows from 10 to 100, corresponding to an increase of approximately 9\%. Memory consumption grows gradually from about 2.7~GB to 2.9~GB over the same range. These results indicate that increasing the number of OLTs entities has a limited impact on simulation cost, suggesting that the simulator scales efficiently with infrastructure size.

Table~\ref{tab:scalability_users} reports the results when varying the number of users while keeping the number of OLTs fixed at 100. In this case, runtime increases from approximately 25~s for 100 users to 157~s for 1{,}000 users. Memory consumption grows from about 1.2~GB to 3.0~GB across the same range. This trend reflects the increased number of simulated entities and workload events generated by user devices. Despite this growth, the simulator remains capable of handling scenarios with up to 1{,}000 users and 100 OLTs within a few minutes of runtime on a standard workstation.

\begin{table}[h]
\centering
\caption{Scalability results when varying the number of OLTs (1{,}000 users).}
\label{tab:scalability_olts}
\small
\begin{tabular}{ccc}
\toprule
\textbf{OLTs} & \textbf{Simulation Time (s)} & \textbf{Max RSS (MB)} \\
\midrule
10  & 142.29 & 2678.3 \\
20  & 144.76 & 2711.6 \\
30  & 143.83 & 2756.2 \\
40  & 146.27 & 2779.7 \\
50  & 145.12 & 2780.3 \\
60  & 147.54 & 2810.6 \\
70  & 148.16 & 2817.9 \\
80  & 146.29 & 2824.9 \\
90  & 151.26 & 2863.6 \\
100 & 154.77 & 2883.5 \\
\bottomrule
\end{tabular}
\end{table}

\begin{table}[h]
\centering
\caption{Scalability results when varying the number of users (100 OLTs).}
\label{tab:scalability_users}
\small
\begin{tabular}{ccc}
\toprule
\textbf{Users} & \textbf{Simulation Time (s)} & \textbf{Max RSS (MB)} \\
\midrule
100  & 25.41  & 1224.3 \\
200  & 35.49  & 1473.0 \\
300  & 48.54  & 1707.4 \\
400  & 62.47  & 1971.6 \\
500  & 75.53  & 2225.0 \\
600  & 90.33  & 2472.3 \\
700  & 104.47 & 2597.5 \\
800  & 126.15 & 2789.9 \\
900  & 141.53 & 2856.8 \\
1000 & 156.52 & 2960.5 \\
\bottomrule
\end{tabular}
\end{table}

Overall, these results show that \toolname{} scales well with both infrastructure size and number of simulated users. Runtime grows approximately linearly with the number of workload-generating devices, while increases in infrastructure elements (OLTs) have a relatively small impact on simulation cost. These findings confirm that the simulator can support large-scale PON-enabled edge scenarios with hundreds of nodes and thousands of users while maintaining practical execution times and memory usage.

\section{Discussion}
\label{sec:discussion}
While the experiments presented in Section~\ref{sec:exp} illustrate two representative use cases of \toolname{}, namely capacity planning and orchestration policy evaluation, it is important to discuss the realism, assumptions, and intended scope of the simulator.
 
\paragraph{Simulation fidelity and scope}
\toolname{} adopts a system-level simulation approach aimed at evaluating orchestration strategies and resource provisioning decisions in PON-enabled edge infrastructures. Rather than reproducing detailed protocol behavior, the simulator captures the architectural characteristics that most strongly influence orchestration outcomes, including hierarchical infrastructure organization (ONTs, OLTs, and cloud nodes), heterogeneous compute capacities across tiers, hybrid VM-container execution environments, and the separation between service placement and runtime task offloading. Together with configurable network latency and bandwidth parameters and heterogeneous user workloads, there elements allow the simulator to capture the key system-level effects such as resources contention, queueing delays, and communication latency across the edge hierarchy.
To maintain scalability and support large multi-tier deployments, some aspects or real systems are intentionally abstracted. In particular, \toolname{} does not model packet-level PON mechanisms (e.g., dynamic bandwidth allocation or optical framing) and represents network communication through latency and bandwidth abstractions. Likewise, container and VM execution are modeled through resource consumption and scheduling delays rather than detailed emulation of container engines, hypervisor internals, or orchestration platforms such as Kubernetes. Control-plane functions such as monitoring, logging, and cluster management are therefore not explicitly simulated.

In addition, the simulator is calibrated using representative hardware configurations and workload characteristics derived from prior studies~\cite{cloudsim, mechalikh2019pureedgesim} and industrial input from the GENIO project~\cite{cesarano2025genio}. While \toolname{} is not calibrated against a full-scale experimental deployment, the chosen parameters ensure that simulated scenarios remain within realistic operating ranges. Consequently, the results should be interpreted as system-level insights rather than precise performance predictions. In particular, the simulator is well suited for comparative evaluation of orchestration policies and for capacity planning studies, whereas protocol-level performance validation would require complementary experimental approaches. 

\paragraph{Extensibility of the simulator}
The placement and offloading strategies evaluated in this work are heuristic policies intended to serve as representative baselines. In particular, the implemented algorithms capture different classes of orchestration strategies commonly studied in edge computing environments, including simple load-balancing policies (e.g., round-robin), resource-aware placement heuristics, and delay-aware offloading decisions.
At the same time, \toolname{} is designed as an extensible simulation environment for evaluating alternative orchestration mechanisms. The modular architecture of the \texttt{ContainerOrchestrator} and \texttt{TaskOrchestrator} components allows new placement and offloading policies to be integrated without modifying the rest of the simulation framework. This design enables experimentation with more advanced approaches such as optimization-based and learning-based policies. 

\paragraph{Energy Modeling}
\toolname{} includes support for energy modeling at both the network and infrastructure levels. In particular, the simulator allows nodes and network links to be associated with configurable energy consumption parameters, enabling the study of energy-aware resource management strategies. In this work, however, the experimental analysis focuses on two specific use cases: capacity planning and orchestration policy comparison under heterogeneous workloads. For this reason, the evaluation concentrates on performance-oriented metrics such as task success rate and end-to-end latency. Energy-aware evaluation is an important direction for future work.

\section{Conclusion}
\label{sec:conclusion}
We presented \toolname{}, a simulation platform to support PON-enabled edge infrastructures. Simulation scenarios inspired by the GENIO project showed that far-edge ONTs reduce latency only under moderate requirements, while strict workloads demand higher provisioning at the OLT. In multi-OLT setups, orchestration proved decisive: advanced placement and delay-aware offloading strategies consistently outperformed simple heuristics, particularly under bursty applications such as streaming and gaming. Beyond the presented evaluation, \toolname{} can also serve as a foundation for future research, enabling the assessment of novel orchestration policies and adaptive scheduling mechanisms in heterogeneous PON–edge environments. Moreover, industry practitioners can employ the tool to analyze how alternative orchestration strategies perform for their specific application scenarios, supporting data-driven decisions in deployment planning. Overall, the experimental analysis demonstrates the feasibility of combining PON networks with edge computing workloads by provisioning sufficient hardware resources within the PON infrastructure and adopting appropriate orchestration algorithms.

\section*{Acknowledgment}
This work has been partially supported by the GENIO project (CUP B69J23005770005) funded by MIMIT, "Accordi per l'Innovazione" program. We also gratefully acknowledge the valuable contribution of Master’s student Gianluigi Erra.

\balance

\bibliographystyle{IEEEtran}
\bibliography{bibliography}

\end{document}